%% file: main.tex
\documentclass[sigconf, nonacm]{acmart}

\setcopyright{none}

\input{mydefs}

\begin{document}
\fancyhead{}

\title{ \method: Differentially Private Streaming Graph Synthesis \\ by Considering Temporal Dynamics}

\author{Quan Yuan}
\affiliation{
  \institution{Zhejiang University}
}
\email{yq21@zju.edu.cn}

\author{Zhikun Zhang}
\affiliation{
  \institution{Zhejiang University}
}
\email{zhikun@zju.edu.cn}

\author{Linkang Du}
\affiliation{
  \institution{Xi'an Jiaotong University}
}
\email{linkangd@gmail.com}

\author{Min Chen}
\affiliation{
  \institution{Vrije Universiteit Amsterdam}
}
\email{m.chen2@vu.nl}

\author{Mingyang Sun}
\affiliation{
  \institution{Peking University}
}
\email{smy@pku.edu.cn}

\author{Yunjun Gao}
\affiliation{%
  \institution{Zhejiang University}
}
\email{gaoyj@zju.edu.cn}

\author{Michael Backes}
\affiliation{
  \institution{CISPA Helmholtz Center for Information Security}
}
\email{director@cispa.de}

\author{Shibo He}
\affiliation{
  \institution{Zhejiang University}
}
\email{s18he@zju.edu.cn}

\author{Jiming Chen}
\affiliation{
  \institution{Zhejiang University \& Hangzhou Dianzi University}
}
\email{cjm@zju.edu.cn}

\begin{abstract}
Streaming graphs are ubiquitous in daily life, 
such as evolving social networks and dynamic communication systems.
Due to the sensitive information contained in the graph, directly sharing the streaming graphs poses significant privacy risks.
Differential privacy, offering strict theoretical guarantees, has emerged as a standard approach for private graph data synthesis.
However, existing methods predominantly focus on static graph publishing, neglecting the intrinsic relationship between adjacent graphs, thereby resulting in limited performance in streaming data publishing scenarios. 
To address this gap, we propose \method, the first differentially private streaming graph synthesis framework that integrates temporal dynamics. 
\method adaptively adjusts the privacy budget allocation mechanism by analyzing the variations in the current graph compared to the previous one for conserving the privacy budget. 
Moreover, \method aggregates information across various timestamps and adopts crucial post-processing techniques to enhance the synthetic streaming graphs.  
We conduct extensive experiments on four real-world datasets under five commonly used metrics.
The experimental results demonstrate the superiority of our proposed~\method.
\end{abstract}

\maketitle

\section{Introduction}
\label{sec:introduction}

Streaming graphs,
composed of a series of continuous graphs, are prevalent in real-world systems such as dynamic social networks~\cite{sekara2016fundamental} and ongoing communication topologies~\cite{khrabrov2010discovering}, among others. 
Analyzing these data allows systems to offer more valuable services, including hotspot tracking~\cite{yu2013anomalous}, behavior analysis~\cite{rossi2013modeling}, and advertising recommendations~\cite{yin2023next}. 
For example, event propagation can be predicted through the analysis of dynamic social networks~\cite{khafaei2019tracing}. 
However, these graph data often contain personal privacy information, such as user relationships and purchase records, which hinders the sharing of these data.

\textit{Differential privacy} (DP)~\cite{dwork2006calibrating}, regarded as the gold standard in the privacy research 
community, has been applied to safeguard the privacy of graph data~\cite{sala2011pygmalion}. 
The core idea of DP is to ensure that the presence or absence of a single node or edge has a limited impact on the final output. 
Most previous works on differentially private graph analysis aim to develop specialized methods for specific tasks, like degree distribution~\cite{hay2009accurate} and clustering coefficient~\cite{wang2013learning}. 
However, these task-specific solutions are not versatile enough to accommodate arbitrary downstream 
tasks, limiting their applicability.

In this paper, we focus on a more universal and realistic paradigm: publishing synthetic streaming graphs composed of continuously evolving graphs while satisfying DP, thereby enabling various downstreaming graph analysis tasks. 
There are several existing studies investigating \textit{static graph synthesis} with DP guarantee~\cite{chen2014correlated,nguyen2015differentially,xiao2014differentially,yuan2023privgraph}.
These static graph synthesis methods typically involve encoding the original graph, perturbing the encoded information, and then reconstructing the entire graph or directly perturbing the graph's adjacency matrix.
The most straightforward method for \textit{streaming graph synthesis} is to evenly allocate the total privacy budget and independently publish the graph at each timestamp using existing static synthesis methods.

However, the edge changes in streaming graphs often exhibit certain dynamic pattern, with changes in adjacent timestamps generally being relatively gentle.
The above approach fails to utilize this temporal dynamics between adjacent consecutive graphs, 
resulting in a waste of privacy budget and low fidelity of the generated stream data.
Currently, there is a lack of differentially private synthesis method designed based on the temporal dynamics of streaming graphs.

\mypara{Our Proposal}
We propose \method, which leverages the temporal dynamics of 
the 
streaming graphs across different timestamps for more efficient privacy budget allocation and enhanced utility. 
The core idea is to adjust the allocation of the privacy budget according to the severity of the streaming graph changes.
Considering the characteristics of streaming data, we develop two distinct privacy budget allocation strategies. 
The appropriate allocation method is selected based on the dynamic changes observed in the streaming graphs between consecutive timestamps. 
For significant changes, we allocate a portion of the total privacy budget to effectively capture these variations. 
In contrast, for minor variations, the privacy budget for capturing information can be retained, and more budget is allocated to information perturbation in order to enhance synthetic data utility.

More specifically, to preserve key information while reducing disruption, \method first employs communities (\ie, the sets of highly interconnected nodes)~\cite{blondel2008fast} as granularity to extract and perturb graph data. 
Recognizing that the community structure of streaming graphs can undergo subtle or drastic changes at continuous timestamps, which can be captured for better synthetic quality,
\method adopts a community judgment mechanism 
to deal with the variations in the first phase.
If changes between adjacent timestamps are small, \method chooses to retain the previous community partition to conserve privacy budget. 
Conversely, \method tries to re-partition all nodes if there are significant variations.
During the next phase,
\method obtains the noisy node degree information and edge counts between communities.
If the community partition of the current timestamp is from the previous timestamp, 
\method tries to integrate the node information of the two timestamps.
Based on the perturbed information, 
in the last phase,
\method reconstructs the entire graph within and between communities using different edge probability formulas.
Necessary post-processing is utilized to enhance the fidelity of the generated streaming graphs.

\mypara{Evaluation}
We conduct experiments on four real-world streaming graph datasets to demonstrate the superiority of \method. 
The experimental results show that \method outperforms other solutions in most cases. 
For instance, with a privacy budget of 1, \method achieves a 143.5\% lower KL divergence in degree distribution compared to the best baseline method
on the Cit-HepPh dataset.
We further analyze the effect of various components in \method through ablation studies and explore the impact of hyperparameter settings.
The related results highlight the crucial role of temporal dynamics in enhancing the performance of \method.
\rev{For example, when the privacy budget is 2, \method leveraging temporal dynamics achieves an 85.1\% higher overlap ratio of eigenvalue nodes on the Reddit dataset compared to \method that ignores this feature (\ie, independently perturbation at each timestamp). } 
We also explore the scalability of different methods
and find that \method performs better than most baselines.
We further evaluate the performance of all methods across various timestamps,
and observe that \method almost outperforms other competitors across all timestamps.

\mypara{Contributions}
In summary, the main contributions of the paper are three-fold:

\begin{itemize}
[itemsep=2pt,topsep=2pt,parsep=0pt]

\item We propose \method, the first differentially private graph synthesis framework designed for streaming graphs.

\item \method employs communities as granularity to extract and perturb information from the original graph, reducing noise injection while preserving critical details.
Furthermore,
\method leverages temporal dynamics across various timestamps in stream data, introduces a community judgment mechanism to conserve privacy budgets, and incorporates post-processing to enhance the utility of synthesized streaming graphs.

\item We conduct %
experiments on multiple datasets and metrics to illustrate the effectiveness of \method.

\end{itemize}

\section{Preliminaries}
\label{sec:preliminary}

\subsection{Differential Privacy}
\label{subsec:differential privacy}

Differential Privacy (DP)~\cite{dwork2006calibrating} was originated for the data privacy protection scenarios, where a trusted data curator collects data from individual users, perturbs the aggregated results, and then publishes them.
Intuitively, DP guarantees that any single sample from the dataset has a limited impact on the output. 

\begin{definition}[$\varepsilon$-Differential Privacy]
{An algorithm $\mathcal{A}$ satisfies $\varepsilon$-differential privacy ($\varepsilon$-DP), where $\varepsilon>0$, if and only if for any two neighboring datasets $D$ and $D^{\prime}$, and for any possible output set $O$, we have} 
\begin{equation*}
    \Pr {\mathcal{A}(D) \in O }
    \le e^{\varepsilon} \Pr {\mathcal{A}(D^{\prime}) \in O} 
\end{equation*}
\end{definition}

Here, we consider two datasets $D$ and $D^{\prime}$ to be \emph{neighbors}, denoted as $D\simeq D^{\prime}$, if and only if $D = D^{\prime} + r $ or $ D^{\prime} = D + r$, where $D + r$ represents the dataset resulted from
adding the record $r$ to 
$D$.

\mypara{Laplace Mechanism}
This mechanism 
achieves DP requirements by adding random Laplace noise to the aggregated results.
The magnitude of the added noise depends on ${GS}_f$, \ie, the \emph{global sensitivity}, %
\begin{equation*}
{GS}_f = \max_{D\simeq  D^{\prime} } {\parallel f(D) - f(D^{\prime}) \parallel }_1, 
\end{equation*}
where $f$ is the aggregation function. 
When $f$ outputs a scalar, the Laplace mechanism $\mathcal{A}$ can be given below:
\begin{equation*}
    \mathcal{A}_f(D) = f(D) + {Lap}\left(\frac{{GS}_f}{\varepsilon}\right), 
\end{equation*}
where ${Lap}(\beta)$ denotes a random variable sampled from the Laplace distribution $\Pr {{Lap}(\beta)=x}=\frac{1}{2\beta}e^{-\left | x \right | / \beta}$.

\mypara{Composition Properties of DP}
The following composition properties of DP are commonly adopted for building complex differentially private algorithms from simpler subroutines~\cite{dwork2014algorithmic}. 

\begin{itemize}
    \item \mypara{Sequential Composition}
    Combining multiple subroutines that satisfy differential privacy for
    $\{\varepsilon_{1}, \cdots,\varepsilon_{k}\}$ leads to a mechanism satisfying
    $\varepsilon$-DP for $\varepsilon = { \sum_{i}\varepsilon_{i}} $.
    
    \item \mypara{Parallel Composition}
    Given $k$ algorithms working on disjoint subsets, 
    each satisfying DP for $\{\varepsilon_{1},\cdots,\varepsilon_{k}\}$, the result satisfies $\varepsilon$-DP for $\varepsilon=\max\{\varepsilon_{i}\}$.
    
    \item \mypara{Post-processing}
    Given an $\varepsilon$-DP algorithm $\mathcal{A}$, releasing $g(\mathcal{A}(D))$ for any $g$ still satisfies $\varepsilon$-DP. In other words, post-processing an output of a differential private algorithm does not result in additional loss of privacy. 
\end{itemize}

\subsection{Differentially Private Stream Analysis}
\label{subsec:dp_stream}
To address the privacy concerns in the release of  the data stream, 
\emph{event-level} privacy and \emph{user-level} privacy are first proposed~\cite{dwork2010differential}. 
Event-level privacy safeguards a single timestamp within a data stream, which may not offer sufficient privacy guarantees for an entire data stream. 
On the other hand, user-level privacy is dedicated to concealing all timestamps in a stream, providing stronger privacy assurances. 
However, it is inappropriate for infinite streams, which necessitates an infinite amount of perturbation. 
To balance privacy protection and data utility, 
\emph{w-event privacy} is proposed to safeguard arbitrary $w$ consecutive timestamps in a stream~\cite{kellaris2014differentially}. 

We first introduce the notion about \emph{stream prefix} and \emph{neighboring streams}.
Let $T=\{D_1,D_2,\cdots\}$ represents an infinite series, 
and $T[i]=D_i$ corresponds to the data at $i$-th time step. 
The stream prefix $S_t$ is defined as $T_t = \{D_1,D_2,\cdots,D_t\}$, representing the sequence of values up to $t$-th time step. 
Based on this, the concept of $w$-neighboring can be given as follows: 

\begin{definition}[$w$-neighboring]
Two stream prefixes $T_t$ and ${T}_t^{\prime}$ are $w$-neighboring,
if for each pair of ${T_t}[i]$, ${T_t^{\prime}}[i]$ such that $i\in \{1,2, \cdots, t\}$ and $T_t[i]\ne {T_t^{\prime}}[i]$, it holds that $T_t[i]$, ${T_t^{\prime}}[i]$ are neighboring; and for each $T_t[i_1], T_t[i_2], {T_t^{\prime}}[i_1], {T_t^{\prime}}[i_2]$ with $i_1 \le i_2, T_t[i_1] \ne {T_t^{\prime}}[i_1]$ and $T_t[i_2] \ne {T_t^{\prime}}[i_2]$, it holds that $i_2-i_1+1\le w$.
\end{definition}
Two stream prefixes are $w$-neighboring, which means that all their pairwise unequal values could fit in a window of up to $w$ timestamps. 

\begin{definition}[$w$-event privacy]
{An algorithm $\mathcal{A}$ satisfies $w$-event $\varepsilon$-DP (or $w$-event privacy), where $\varepsilon>0$, if and only if for any two $w$-neighboring stream prefixes $T_t$ and $T_{t}^{\prime}$, and for any possible output set $O$, we have} 
\begin{equation*}
    \Pr {\mathcal{A}(T_t) \in O }
    \le e^{\varepsilon} \Pr {\mathcal{A}(T_t^{\prime}) \in O} 
\end{equation*}
\end{definition}

A protection mechanism satisfies $w$-event privacy can provide $\varepsilon$-DP guarantee in any sliding window of size $w$. 
If $w=1$, it degenerates to event-level privacy. 
When $w$ is set to the length of a finite stream, $w$-event privacy converges to user-level privacy.

\subsection{Differentially Private Graph Analysis}
\label{subsec:dp_graph}
Edge-DP~\cite{hay2009accurate} provides rigorous theoretical guarantees to protect the privacy of the edges by limiting the impact of any edges in the graph on the output. 
Specifically, given a graph $G=(V,E)$, its edge neighboring graph $G^{\prime}=(V^{\prime},E^{\prime})$ is originated by adding (or removing) an edge, where $V$ ($V^{\prime}$) denotes the set of nodes and $E$ ($E^{\prime}$) denotes the set of edges. 
According to~\cite{hay2009accurate},
the system difference $a \oplus b$ is the sets of elements in either set $a$ or set $b$, but not in both, \ie, $a \oplus b = (a \cup b) \setminus (a \cap b)$. 
Therefore, the definitions of edge neighboring graph and $\varepsilon$-edge DP are as follows.

\begin{definition}[Edge neighboring graph]
Given a graph $G=(V,E)$, a graph $G^{\prime}=(V^{\prime},E^{\prime})$ is an edge neighboring graph of $G$ if and only if $\left | V \oplus V^{\prime} \right | + \left | E \oplus E^{\prime} \right | = 1 $.
\end{definition}

\begin{definition}[$\varepsilon$-edge differential privacy]

An algorithm $\mathcal{A}$ satisfies $\varepsilon$-edge differential privacy ($\varepsilon$-edge DP), where $\varepsilon>0$. If and only if for any two edge neighboring graphs G and $G^{\prime}$, and for any possible output set $O$, we have
\begin{equation*}
 \Pr {\mathcal{A}(G) \in T}\le e^{\varepsilon}  \Pr {\mathcal{A}(G^{\prime}) \in T} 
\end{equation*}
\end{definition}

According to the privacy definitions of stream data and graph data,
we can further give the concepts of $w$-edge neighboring graph and $w$-event edge privacy.

\begin{definition}[$w$-edge neighboring graph]
Two streaming graph prefixes $G_t$ and ${G}_t^{\prime}$ are $w$-edge neighboring,
if for each ${G_t}[i],{G_t^{\prime}}[i]$ such that $i\in \{1, 2, \cdots, t\}$ and
$G_t[i]\ne {G_t^{\prime}}[i]$, it holds that $G_t[i], {G_t^{\prime}}[i]$ are edge neighboring graphs; and for each $G_t[i_1], G_t[i_2], {G_t^{\prime}}[i_1], {G_t^{\prime}}[i_2]$ with $i_1 \le i_2, G_t[i_1] \ne {G_t^{\prime}}[i_1]$ and $G_t[i_2] \ne {G_t^{\prime}}[i_2]$, it holds that $i_2-i_1+1\le w$.
\end{definition}

\begin{definition}[$w$-event edge privacy]
{An algorithm $\mathcal{A}$ satisfies $w$-event $\varepsilon$-edge DP (or $w$-event edge privacy), where $\varepsilon>0$, if and only if for any two $w$-edge neighboring streaming graph prefixes $G_t$ and $G_{t}^{\prime}$, and for any possible output set $O$, we have} 
\begin{equation*}
    \Pr {\mathcal{A}(G_t) \in O }
    \le e^{\varepsilon} \Pr {\mathcal{A}(G_t^{\prime}) \in O}
\end{equation*}
\end{definition}

$w$-event edge privacy can achieve edge protection for any continuous $w$ graphs with a privacy budget of $\varepsilon$ in the streaming graphs.

\section{Problem Definition and Strawman Solution}
\label{sec:problem_definition}

\subsection{Problem Definition}
\label{subsec:problem_definition}
There are many scenarios in real life that require analysis of continuously changing graph data to provide services, \eg, social networks and transportation networks. 
Due to privacy concerns, the trusted curator usually publishes synthetic data that satisfies differential privacy instead of directly releasing the original data~\cite{zhu2017differentially}. 
The synthetic graph dataset is expected to retain similar properties to the original graph and serve as a secure substitute for the original data. 
In this paper, we consider an undirected and unweighted graph database $\mathcal{G}_{orig}$ that consists of streaming graph $G_t=(J_t,E_t)$, where $J_t$ is the set of nodes and $E_t$ is the set of edges at timestamp $t$.
The goal is to find a release mechanism $\mathcal{R}$, which takes inputs from $\mathcal{G}_{orig}$ and outputs a dynamic synthetic graph database $\mathcal{G}_{syn}$ at each $t$.
$\mathcal{R}$ satisfies $w$-event edge privacy.
We summarize the frequently used mathematical notations in \autoref{table:math_notations}.

\begin{table}[!t]
    \centering
    \caption{Summary of mathematical notations.}
    \label{table:math_notations}
    \vspace{-0.3cm}
    \footnotesize
    \setlength{\tabcolsep}{1.2em}
	\begin{tabular}{cc}
		\toprule
		\textbf{Notation} & \textbf{Description}  \\
		\midrule
            $t$ & Timestamp \\
		$G_t$ & Graph at the timestamp $t$ \\
		$\varepsilon$ & Privacy budget  \\
             $w$ & The window size \\
		$N^t$ & The number of nodes of $G_t$ \\
		${m^t}$ & The number of edges of $G_t$  \\
		$\mathbb{C}$ &  The set of community partitions \\
		$C$ & The community composed of nodes \\
         $J$ & The set of all nodes \\
		$D_i$& Degree information within the community \\
            $D_o$& Degree information outside the community \\
		$V$& Edge count between communities  \\
		\bottomrule
	\end{tabular}
\end{table}

\subsection{Strawman Solution}
\label{section:existing_solutions}

A straightforward approach to address this streaming graph publishing problem is to evenly distribute the total privacy budget at each timestamp and then utilize existing differentially private synthesis methods for releasing static graphs. 
These synthesis methods for static graphs typically fall into two categories. 
The first way involves encoding the original static graph, adding noise to the encoded data to satisfy DP, and then reconstructing the graph. 
Another way is to directly perturb the entire adjacency matrix. 

However, streaming graph data exhibits dynamic variations where adjacent graphs can show close relationships or undergo significant changes. 
The current graph may experience minor changes from the previous timestamp, involving only a few edges, or it may undergo significant variations involving a large number of edges.
Accurately capturing these dynamic changes while satisfying the DP guarantee presents considerable challenges for existing methods. 
First, existing static methods are optimized for single graph data publishing problems, thus struggling to effectively handle such dynamics. 
Moreover, the release of streaming data imposes demands on real-time performance, necessitating high efficiency for the designed method. 
Third, in streaming data scenarios, the entire privacy budget needs to be divided for a series of graph data. 
Thus, it is more challenging to achieve a proper balance between privacy protection and data utility.

\section{Our Proposal: \method}
\label{sec:methodology}

\subsection{Motivation}
\label{subsec:motivation}
To address the aforementioned challenges, we develop our approach from two perspectives to ensure data utility, \ie, the fidelity of individual graph data and the dynamics of a series of graph data in a data stream. 
For each individual graph data in the data stream, we want to retain as much information as possible, necessitating careful encoding of graph data. 
When synthesizing a graph under edge-DP, directly perturbing each edge in the adjacency matrix may introduce excessive noise, whereas compressing the graph into node degrees can lead to significant information loss. 
According to~\cite{yuan2023privgraph}, community (\ie, the set of closely connected nodes) is a suitable granularity that can reduce noise while preserving key information.
Inspired by this, we choose to partition all nodes into communities and perturb the aggregated information. 
For the dynamics of a series of graph data, we aim to identify and leverage the variations of streaming graphs in privacy-preserving synthesis.
We hope to select an appropriate privacy budget allocation strategy according to the extent of variations between adjacent timestamps
in order to enhance the utility of the data stream. 
Based on the above considerations, we design a streaming graph publishing method, called \method, which rigorously adheres to $w$-event privacy requirements while ensuring high data utility.

\subsection{Overview}
\begin{figure*}[htbp]
\centering
\includegraphics[width=0.92\textwidth] {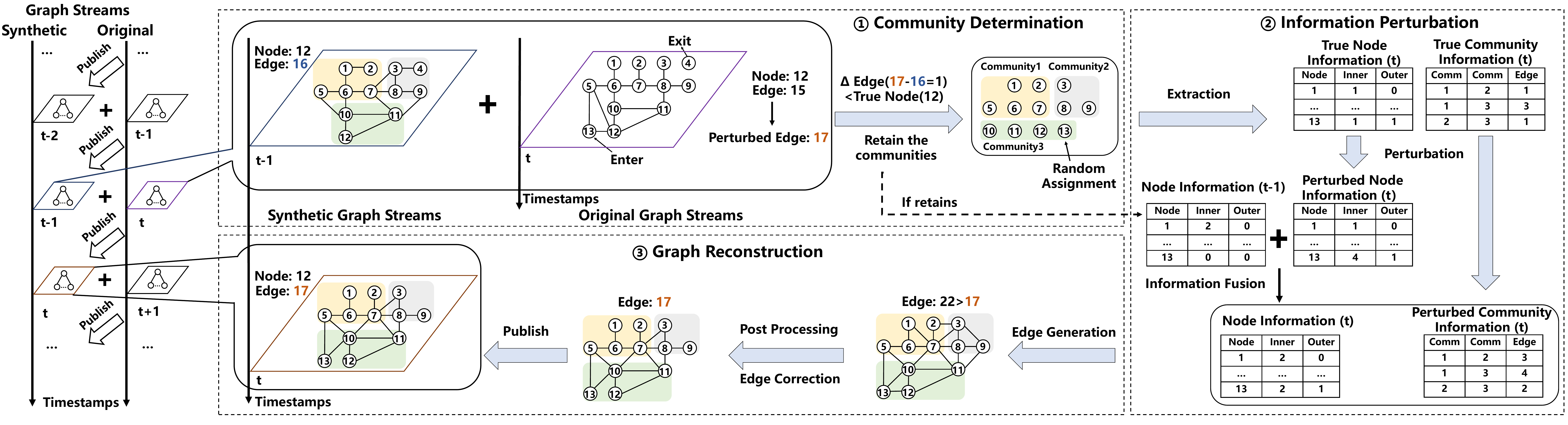}
\vspace{-0.3cm}
\caption{\method overview. \method is composed of three phases: Community determination, information perturbation, and graph reconstruction.
In the community determination phase, the difference in edges between the last and current graphs is adopted to determine whether to retain the previous community division or community re-partitioning. 
In the information perturbation phase, node degree information within and across communities, as well as inter-community connections, is extracted and perturbed to satisfy DP. 
If the current community partition is from the last partition, the final estimated result integrates the perturbed information from both the current and last timestamps.
An initial graph is constructed using the edge probability formulas in the graph reconstruction phase. 
Then, \method utilizes post-processing to adjust the generated graph.
}
\vspace{-0.15cm}
\label{fig:framework}
\end{figure*}

As depicted in \autoref{fig:framework}, the workflow of \method mainly consists of three phases: Community determination, information perturbation, and graph reconstruction. 
The new graphs are continually generated through the three phases, forming streaming graphs. 
Based on the original graph dataset, the trusted aggregator maintains a synthetic dataset for publishing. 

\mypara{Phase 1: Community Determination}
We design a dynamic community determination mechanism to obtain node partitioning for each timestamp. 
The core idea is to determine the community partitioning strategy based on the differences between the current and previous graphs. 
If the variations in these graphs are small, we can stick with the previous community divisions, thereby conserving the privacy budget for the following phases. 
Conversely, if there are significant changes over time, we need to re-partition the nodes in the graph. 
The details of Phase 1 can be found in~\autoref{subsec:community_determination}. 

\mypara{Phase 2: Information Perturbation}
Based on the community partitioning results from Phase 1, we further encode and perturb three kinds of edge information, \ie, the node's connections in its community, the node's connections to other communities, and the connections between different communities. 
To enhance the accuracy of the data, we choose to integrate the perturbed information from both the current timestamp and the previous timestamp if the variations between the graphs are not significant. 
The details of Phase 2 are in~\autoref{subsec:information_pert}. 

\mypara{Phase 3: Graph Reconstruction}
Using the perturbed information obtained in the second phase, \method reconstructs the edges both within and between communities according to different probabilistic edge connection formulas. 
To maintain a more realistic graph distribution, \method performs post-processing on the initially generated graph, ensuring that the total number of edges in the synthetic graph is close to the original graph.
The details of Phase 3 are referred to~\autoref{subsec:graph_recons}.

\subsection{Community Determination}
\label{subsec:community_determination}

During the community determination phase, it is crucial to assess the changes between the current and previous graphs. 
Here, we judge an obvious change between the two graphs by examining the variation in the number of edges. 
The reasons are as follows.
Firstly, the evolution of the community structure is highly correlated with the dynamic changes of edges in the graph~\cite{leskovec2008statistical}. 
Concretely, the core feature of a community is that its internal connections are dense and its external connections are sparse, and the dynamic changes of edges directly impact this structure.
Secondly, the edge count in a graph tends to be a large value, and its sensitivity is only 1, allowing it to accurately reflect the feature of the original graph under a small privacy budget. 
Additionally, the number of edges can assist the graph reconstruction process detailed in~\autoref{subsec:graph_recons}.

\autoref{algorithm:community_determine} outlines the basic process of community determination.
We first apply the Laplace mechanism at each moment to obtain the number of perturbed edges for subsequent judgment and processing.
The privacy budget consumed in this step is $\varepsilon_e$, with sensitivity $\Delta f_e=1$.
When synthesizing the first graph in the streaming data, historical information is unavailable, necessitating the allocation of a privacy budget for community partitioning. 
Here, we utilize the community division method proposed in~\cite{yuan2023privgraph} to achieve differentially private community partitioning (\ie, $Comm\_{div}$).
The general idea is to first randomly merge the nodes of the original graph into several super-nodes. 
Subsequently, it employs the Louvain approach~\cite{blondel2008fast} to partition the perturbed super-node graph into communities. 
Finally, considering the number of edges of the nodes in various communities, the exponent mechanism~\cite{mcsherry2007mechanism} is applied to fine-tune the community assignments of the individual nodes.
Following the aforementioned operations, we can obtain the community partitioning result $\mathbb{C_D}$ for the initial graph. 

When $t>1$, the previous graph information can aid in determining the community structure of the current graph. 
First, the change in edge count $\delta_e$ between the two graphs can be estimated based on the number of perturbed edges in the previous and current steps. 
According to~\cite{leskovec2007graph},
many graphs densify over time, with the number of edges growing super-linearly in the number of nodes. 
Thus, we choose the number of nodes in the graph as a basis to judge whether there has been a significant difference between the communities of the two graphs.  
If $\delta_e$ exceeds the number of nodes $N^t$, indicating a significant alteration in the graph's community structure, conducting a new node partition is necessary. 
This process is similar to the community partitioning method adopted at the first timestamp. 

On the other hand, if the variation in the number of edges is small, the community partitioning from the previous step can be reused. 
In this case, the surplus privacy budget $\varepsilon_c^t$ is reserved for the subsequent phase of information perturbation, which helps to reduce the intensity of the perturbation.

\begin{algorithm}[!t]

        \caption{Community Determination}
        \label{algorithm:community_determine}
        \KwIn{Current timestamp $t$, original graph $G_t$, privacy budget $\varepsilon_{e}^{t}$, $\varepsilon_{c}^t$, the number of nodes of current graph $N^t$, the number of perturbed edges of last graph $m_{pert}^{t-1}$ (if $t>1$),
        the last community division $\mathbb{C_{L}}$ (if $t>1$)
        }
        \KwOut{the number of perturbed edge $m_{pert}^t$, community division $\mathbb{C_{D}}=\{C_1^t,C_2^t,\dots\}$}

        Obtain $m_{pert}^t$ by perturbing the true value $m^t$  \\
        $m_{pert}^{t} \leftarrow m^t + Laplace(\varepsilon_e^t,\Delta f_e)$  \\
        \eIf{$t=1$}
        {
        Perform community partitioning on nodes of $G_t$ \\
        $\mathbb{C_{D}} \leftarrow Comm\_{div}(G_t,\varepsilon_c^t)$
        }{
        $\delta_e=| m_{pert}^t - m_{pert}^{t-1} |$  \\
        // Community judgment \\
            \eIf{$\delta_e > N^t$}
            {
            $\mathbb{C_{D}} \leftarrow Comm\_{div}(G_t,\varepsilon_c^t)$
            }{
            // $\varepsilon_c^t$ is reserved for the subsequent phase\\
            $\mathbb{C_{D}} \leftarrow \mathbb{C_{L}}$ \\
            Update $\mathbb{C_{D}}$ by randomly assigning newly added nodes at time $t$ to existing communities.
            }
        }
       
\end{algorithm}

\subsection{Information Perturbation}
\label{subsec:information_pert}

Based on the community partition $\mathbb{C_D}$ obtained in the first phase, 
information perturbation can be strategically conducted from both node-level and community-level perspectives, thereby achieving a good balance between privacy and utility.
\autoref{algorithm:information_pert} illustrates the process of the information perturbation phase.

The first step is to extract information from nodes and communities.
For each node, we count the number of edges within its community $D_i^t$ and the number of edges that connect to other communities $D_o^t$ separately. 
On the one hand, this operation effectively aggregates the connection information of nodes, avoiding excessive perturbation. 
On the other hand, gathering information under community granularity helps preserve nodes' characteristics. 
Considering that the number of communities is usually large, we also calculate the total number of external edges between any two communities $V^t$. 
This helps to accurately capture edge information across various communities under high privacy protection requirements.
The dimensions of $D_i^t$ and $D_o^t$ are $|J|$, where $|J|$ is the number of nodes.
The dimension of $V^t$ is $\frac{H(H-1)}{2}$,
where $H$ is the number of communities.

In the second step, \method perturbs the information extracted in the first step, \ie, the node degrees and edge counts between communities. 
The Laplace mechanism is applied to satisfy differential privacy. 
Since an edge can influence two nodes simultaneously, the sensitivity of degree information $\Delta f_d$ is 2. 
The sensitivity of edge count between communities $\Delta f_v$ is 1. 
Note that the degree information outside the community and the edge count between different communities overlap (\ie, they both visit the edges between communities), 
necessitating the consumption of separate privacy budgets, $\varepsilon_{i1}^t$ and $\varepsilon_{i2}^t$, respectively.
Here, 
we set $\varepsilon_{i1}^t=\varepsilon_{i2}^t=0.5\varepsilon_{i}^t$ to achieve a balance between degree information and edge count. 
Conversely, degree information for nodes within the community does not overlap with the above parts.
Hence, the degree of nodes within the community can be protected using the budget of $\varepsilon_i^t$.

In practice, the perturbed values of the node's degree or edge count may be negative, which conflicts with their definitions. 
To address this issue, we employ the NormSub~\cite{wang2019consistent} method to process the perturbed values. 
Through this consistency processing, we ensure that the degree information and edge count adhere to non-negative constraints.

In real-world scenarios, many streaming graphs demonstrate relatively consistent features between successive timestamps,
which can aid in obtaining more accurate information.
Here, we assess the necessity of utilizing information from the last timestamp based on the sources of community partition.
If the current community partitioning $\mathbb{C_D}$ builds on previous partitioning results $\mathbb{C_L}$, we can integrate information from both steps to achieve a more precise estimate, especially under limited privacy budgets. 

In particular, we employ a weighted approach to obtain the final degree estimate for overlapping nodes at two timestamps. 
By taking into account the privacy budget of the two parts,
the final degree information can be updated as follows: 

\begin{equation}
\begin{aligned}
    \label{eq:weighted_update}
    \hat{D}_i^t &= \alpha_1 \bar{D}_i^t + (1-\alpha_1)\hat{D}_i^{t-1}, \\
    \hat{D}_o^t &= \alpha_2 \bar{D}_o^t + (1-\alpha_2)\hat{D}_o^{t-1},
\end{aligned}
\end{equation}
where $\alpha_1=\frac{\varepsilon_i^{t}}{\varepsilon_i^{t}+\varepsilon_i^{t-1}}$ and $\alpha_2=\frac{\varepsilon_{i1}^{t}}{\varepsilon_{i1}^{t}+\varepsilon_{i1}^{t-1}}$.
Our rationality is that
the higher the privacy budget, the greater the fidelity of information; therefore, the corresponding weight should be higher. 
By considering the privacy budget and node degree information across adjacent timestamps, we can better capture the dynamic features in the streaming graphs. 
We further verify the effect of this step in~\autoref{subsec:ablation}.
Note that NormSub and Combine only access noise-perturbed data instead of real data, and thus can be regarded as post-processing.

\begin{algorithm}[!t]

        \caption{Information Perturbation}
        \label{algorithm:information_pert}
        \KwIn{Current timestamp $t$, Original graph $G_t$, current community division $\mathbb{C_D}$,  privacy budget $ \varepsilon_i^t, \varepsilon_{i1}^t, \varepsilon_{i2}^t=\varepsilon_i^t - \varepsilon_{i1}^t $
        }
        \KwOut{Noisy degree information within the community $\hat{D}_{i}^{t}$, noisy degree information outside the community $\hat{D}_{o}^{t}$, noisy edge count between various communities $\hat{V}^t$
        }
       
        \textbf{Step 1: Information extraction} \\
        
            \For{$node$ in $G_{t}.nodes()$}
                {
                Extract the node's degree information $d_i$ within its community, and the node's degree information $d_o$ outside its community
                 \\
                Update $D_i^t$ and $D_o^t$ based on $d_i$ and  $d_o$ \\
                
                }
            \For{$C_a,C_b(a \ne b)$ in $\mathbb{C_D}$}
                {
                Extract the number of edges $v_{a,b}$ between community $C_a$ and $C_b$ \\
                Update $V^t$ based on $v_{a,b}$
                }

        \textbf{Step 2: Noise injection} \\
        // The edges within and outside the community are non-overlapping \\
        $\tilde{D}_i^t \leftarrow D_i^t + Laplace(\varepsilon_{i}^t,\Delta f_d)$ \\
        // The degree information outside the community and the edge count between various communities consume the total privacy budget $\varepsilon_i^t$ \\
        $\tilde{D}_o^t \leftarrow D_o^t + Laplace(\varepsilon_{i1}^t,\Delta f_d)$,
        $\tilde{V}^t \leftarrow V^t + Laplace(\varepsilon_{i2}^t,\Delta f_v)$ \\
        // The following steps are post-processing\\
        \textbf{Step 3: Consistency processing} \\
        $\bar{D}_i^t \leftarrow NormSub(\tilde{D}_i^t)$,
        $\bar{D}_o^t \leftarrow NormSub(\tilde{D}_o^t)$,
        $\bar{V}^t \leftarrow NormSub(\tilde{V}^t)$  \\
        
        \textbf{Step 4: Information fusion} \\
        \eIf{$\mathbb{C_D}$ is based on $\mathbb{C_L}$}
        {
        $\hat{D}_i^t \leftarrow Combine(\bar{D}_i^t,\hat{D}_i^{t-1})$,
        $\hat{D}_o^t \leftarrow Combine(\bar{D}_o^t,\hat{D}_o^{t-1})$
        }
        {
        $\hat{D}_i^t \leftarrow \bar{D}_i^t$,
        $\hat{D}_o^t \leftarrow \bar{D}_o^t$
        }

\end{algorithm}

\subsection{Graph Reconstruction}
\label{subsec:graph_recons}

In this phase, we reconstruct the whole graph based on the perturbed degree information and edge count between communities.
Due to the different perturbation strategies of the extracted nodes within and outside the community, 
it is improper to adopt the same reconstruction method for these two types of edges. 
Hence, we initially utilize distinct reconstruction methods for intra-community and inter-community connections and then apply post-processing to the generated graph.  
The detailed process is 
in~\autoref{algorithm:graph_recons}.

For the intra-community edges,
we utilize perturbed $\hat D_i^t$ to generate the subgraph in each community.
Assuming that $\hat{d}$ is the perturbed degree sequence within the community $C_a$, 
the connection probability between node $x$ and node $y$ within $C_a$ is as follows. 

\begin{equation}
    \label{eq:prob_intra_community}
    p_{x \in C_a,y \in C_a} = \frac{\hat d{^{x}_{a}} \cdot \hat d{_{a}^{y}}}{\sum_{z \in C_a} \hat d{^{z}_{a}} }, 
\end{equation}
where $\hat d^x_a$ and $\hat d^y_a$ are the degrees of node $x$ and $y$ within the community $C_a$, 
and the denominator is the sum of node degree in $C_a$. 
Based on \autoref{eq:prob_intra_community}, \method can effectively recover the degree information and satisfy the power-law distribution~\cite{aiello2000random}.

For the connections between communities, the two types of information extracted can be effectively combined, namely the degree information of each node outside the community $\hat D_o^t$ and the number of connections between communities $\hat V^t$.
Assuming that node $x$ belongs to the community $C_a$ and node $y$ belongs to the community $C_b$, our objective is to estimate the probability of connection between nodes $x$ and $y$. 
$\hat h$ represents the perturbed degree sequence outside the community, 
and $\hat v$ represents the perturbed edge count between communities.
Firstly, the expected number of edges from the node to other communities can be calculated. 
For instance, the expected number of edges $\hat e_x^b$ from node $x$ to community $C_b$ is: 
\begin{equation}
    \label{eq:expected_edge}
    \hat e_x^b = \hat h_x \cdot \frac{\hat v_{a,b}}{\sum_{g\in \mathbb{C_D}} \hat v_{a,g}},
\end{equation}
where $\hat h_x$ is the perturbed degree information outside the community $C_a$ for node $x$, 
$\hat v_{a,b}$ is the perturbed edge count between community $C_a$ and community $C_b$,
the denominator is the sum of the number of edges between community $C_a$ and other communities. 
Similarly, we can obtain the expected number of edges $\hat e_y^a$ from node $y$ to community $C_a$.
Based on the expected number of edges mentioned above, we further estimate the probability of connected edges between nodes $x$ and $y$: 
\begin{equation}
    \label{eq:prob_inter_community}
    p_{x \in C_a,y \in C_b} = \frac{\hat e_x^b \cdot \hat e_y^a}{\sum_{z\in C_b} \hat e_z^a},
\end{equation}
where the denominator is the sum of the expected number of edges from all nodes in the community $C_b$ to the community $C_a$. 
Note that the denominator can also be the sum of the expected number of edges from all nodes in the community $C_a$ to the community $C_b$. 

Using the formula mentioned above, we can initialize the connections both within and between communities. 
However, the generated graph may not align perfectly with the original expected information, particularly in cases with significant noise perturbation. 
Thus, we aim to enhance the utility of the generated graph by integrating multiple previously available expected information (\ie, the perturbed total number of edges and the degree information of nodes) for post-processing. 
We try to ensure that the edge count and node degree in the generated graph
closely match those of the original,
as they can well reflect the patterns of the graph.

Firstly, we can obtain the degree information of nodes within and outside the community (\ie, $H_i^t$ and $H_o^t$) and the total number of generated edges $m_g^t$. 
Additionally, for all nodes, we can compute the disparity (\ie $\Delta M_i^t$ and $\Delta M_o^t$) between the degree information (\ie, $\hat D_i^t$ and $\hat D_o^t$) from the information perturbation phase and the generated degree information (\ie, $H_i^t$ and $H_o^t$). 
Furthermore, we merge $\Delta M_i^t$ and $\Delta M_o^t$ into a unified set (\ie, $\Delta M^t$) for subsequent iterations. 
Then, we calculate the difference $\Delta m$ between the perturbed total number of edges $m^t_{pert}$ and the total number of edges $m_g^t$ in the generated graph $G_s^t$, which can guide the termination criteria for post-processing. 
If $\Delta m$ is greater than 0, it indicates that the perturbed number of edges exceeds the total number of edges in the generated graph, necessitating the addition of several edges to $G_s^t$. 
Conversely, if $\Delta m$  is less than 0, it signifies that some edges need to be removed from the generated graph $G_s^t$. 
Next, we illustrate the edge processing steps for $\Delta m > 0$. 
We begin by sorting all relevant nodes in descending order based on $\Delta M^t$. 
A higher order indicates a larger difference, implying that more edges need to be added to $G_{s}^t$. 
Note that there are a total of $2N^t$ nodes in the set $\Delta M^t$, as it contains both $\Delta M_i^t$ and $\Delta M_o^t$. 
We iterate through the sorted nodes. 
In each iteration, we compute the estimated degree information and compare it with the degree information of the generated graph to determine the number of edges $m_u$ that need to be added to node $u$. 
We then randomly add $m_u$ edges to the set of candidate nodes that have not yet been connected. 
Subsequently, we tally the estimated number of edges and the total number of edges in the generated graph and calculate the difference between them $\Delta m_1$. 
If $\Delta m_1 \leq 0$, the total number of edges in the generated graph is approaching the estimated number of edges, prompting the termination of the loop. 
This post-processing effectively adjusts the degree information of the generated graph and aligns its density with that of original data.

\begin{algorithm}[!t]

        \caption{Graph Reconstruction}
        \label{algorithm:graph_recons}
        \KwIn{Null graph $G_N^t$, community division $\mathbb{C_D}$, noisy degree information within the community $\hat D_i^t$, noisy degree information outside the community $\hat D_o^t$, noisy edge count between various communities $\hat V^t$, the number of perturbed edges $m^t_{pert}$ }
        \KwOut{Synthetic graph $G_s^t$}
        Initialize $G_s^t$ as $G_N^t$ \\
        \textbf{Step 1: Intra-community edge generation} \\
        
            \For{$C_a$ in $\mathbb{C_D}$}
                {
                Generate subgraph $S_a^t$ based on $\hat D_{i,a}^t$ and~\autoref{eq:prob_intra_community}
                 \\
                Update $G_s^t$ by $S_a^t$
                
                }

        \textbf{Step 2: Inter-community edge generation} \\
            \For{$C_a,C_b(a \ne b)$ in $\mathbb{C_D}$}
                {
                Generate subgraph $S_{a,b}^t$ based on $\hat V_{a,b}^t$, \autoref{eq:expected_edge} and~\autoref{eq:prob_inter_community}
                \\
                Update $G_s^t$ by $S_{a,b}^t$
                }
        \textbf{Step 3: Edge post-processing} \\
        For the generated graph $G_s^t$,
        calculate the degree information of nodes within the community $H_{i}^t$ and outside the community $H_{o}^t$,
        and the total number of generated edges $m^t_{g}$\\ 
        $\Delta M_i^t \leftarrow \hat D_i^t - H_i^t$,
        $\Delta M_o^t \leftarrow \hat D_o^t - H_o^t$ \\
        $\Delta M^t \leftarrow \{\Delta M_i^t, \Delta M_o^t\}$ \\
        Calculate the difference between the estimated number of edges and generated number of edges \\
        $\Delta m \leftarrow m_{pert}^t - m_g^t$  \\
        \If{$\Delta m > 0$ (or $\Delta m < 0$) }
        {
        Sort $\Delta M^t$ in descending (or ascending) order \\
        \For{node u in sorted $G_s^t$.nodes()}
        {Calculate the number of edges $m_u$ to be added (or removed) for node $u$ \\
        Update $G_s^t$ by randomly adding (or removing) $m_u$ edges from the candidate set \\ 
        Calculate the difference $\Delta m_1$ between $m_{pert}^t$ and the total number of edges in current $G_s^t$ \\
        \If{$\Delta m_1 \leq 0$ (or $\Delta m_1 \geq 0$)}
        {break}
        }
        }
        Obtain the final synthetic graph $G_s^t$
          
\end{algorithm}

\subsection{Putting Things Together}
\label{subsec:method_summary}

The aforementioned three phases form the overall process of proposed~\method, as shown in~\autoref{algorithm:method_summary}.
Based on the total privacy budget $\varepsilon$ and window size $w$,
we can obtain the privacy budget $\varepsilon_s^t$ at each timestamp by average allocation.
The budget allocation for a single timestamp can be adjusted according to the graph. 
Initially, we determine the privacy budget $\varepsilon_e^t$ allocated for perturbing the total number of edges. 
As the total number of edges is usually large and the sensitivity is 1, 
we limit this portion of the privacy budget to a maximum of 0.01.
Furthermore, the remaining privacy budget $\varepsilon_r^t$ is computed (\autoref{line:eps_init}). 
At each timestamp, the perturbed number of edges $m_{pert}^t$ can be obtained (\autoref{line:pert_edge}).
If $t = 1$, the privacy budget $\varepsilon_c^t$ is allocated for community partitioning, and $\varepsilon_i^t$  is allocated for information perturbation (\autoref{line:deal_time_1_begin}-\autoref{line:deal_time_1_end}).
For subsequent timestamps, the information from the last graph can aid in community division and information perturbation at $t$. 
The difference $\delta_e$ in perturbed edges between the adjacent timestamps can be calculated (\autoref{line:cal_delta_e}). 
If $\delta_e$ exceeds the number of nodes, indicating significant changes in the two graphs, the privacy budget $\varepsilon_c^t=0.5\varepsilon_r^t$ is allocated for community re-partitioning (\autoref{line:delta_e_larger_than_n_begin}-\autoref{line:delta_e_larger_than_n_end}). 
Conversely, if the difference is small, the previous community partition can be retained, with $\varepsilon_i^t=\varepsilon_{r}^t$ allocated for information perturbation (\autoref{line:delta_e_smaller_than_n_begin}-\autoref{line:delta_e_smaller_than_n_end}). 
Once the community structure is determined, \autoref{algorithm:information_pert} is employed to perturb true information (\autoref{line:information_pert}). 
Finally, the entire graph is reconstructed using~\autoref{algorithm:graph_recons}.

\begin{algorithm}[!t]

        \caption{\method}
        \label{algorithm:method_summary}
        \KwIn{Current timestamp $t$, original graph $G_t$, total privacy budget $\varepsilon$, window size $w$, 
            privacy budget at current timestamp $\varepsilon_{s}^{t}=\varepsilon / w$, privacy budget for edge perturbation $\varepsilon_e^t$, privacy budget for community division $\varepsilon_{c}^t$, privacy budget for  information perturbation $\varepsilon_i^t$, 
        }
        \KwOut{Synthetic graph $G_{s}^t$ }
        $\varepsilon_e^t=\min(0.01, 0.5\varepsilon_s^t)$, $\varepsilon_r^t=\varepsilon_s^t-\varepsilon_e^t$ \label{line:eps_init} \\
        $m_{pert}^t \leftarrow m^t + Laplace(\varepsilon_e^t, \Delta f_e)$ \label{line:pert_edge}\\
        \eIf{$t=1$}
        {
        \label{line:deal_time_1_begin}
        $\varepsilon_c^t=0.5\varepsilon_r^t$, $\varepsilon_i^t=\varepsilon_r^t-\varepsilon_c^t=0.5\varepsilon_r^t$ \\
        Obtain the community division $\mathbb{C_D}$ using~\autoref{algorithm:community_determine} with $\varepsilon_c^t$\\
        Perturb the true information using~\autoref{algorithm:information_pert} and $\mathbb{C_D}$ with $\varepsilon_i^t$ \label{line:deal_time_1_end}
        }{
        $\delta_e=| m_{pert}^t - m_{pert}^{t-1} |$ \label{line:cal_delta_e}  \\
            \eIf{$\delta_e > N^t $}
            {
            \label{line:delta_e_larger_than_n_begin}
            $\varepsilon_c^t=0.5\varepsilon_r^t$, $\varepsilon_i^t=\varepsilon_r^t-\varepsilon_c^t=0.5\varepsilon_r^t$ \\
            Obtain $\mathbb{C_D}$ by re-partitioning the community using~\autoref{algorithm:community_determine} with $\varepsilon_c^t$  \label{line:delta_e_larger_than_n_end}\\
            }{
            \label{line:delta_e_smaller_than_n_begin}
            $\varepsilon_c^t=0$, $\varepsilon_i^t=\varepsilon_r^t$ \\
            $\mathbb{C_{D}} \leftarrow \mathbb{C_{L}}$ \label{line:delta_e_smaller_than_n_end}\\
            }
            Perturb the true information using~\autoref{algorithm:information_pert} and $\mathbb{C_D}$ with $\varepsilon_i^t$ \label{line:information_pert}
        }
        Reconstruct the graph using~\autoref{algorithm:graph_recons} \label{line:graph_reconstruction}
       
\end{algorithm}

\subsection{Algorithm Analysis}

\mypara{Privacy Budget Analysis}
Recalling \autoref{fig:framework}, 
in each timestamp,
\method includes three phases, \ie, community determination, information perturbation, and graph reconstruction. 
The community determination phase obtains the number of perturbed edges and finishes the community division, 
which consume privacy budgets of $\varepsilon_e^t$ and $\varepsilon_c^t$ respectively. 
The privacy budget consumed by the information perturbation phase is $\varepsilon_i^t$. 
The phase of graph reconstruction does not touch the true data, \ie, consuming no privacy budget. 
Therefore, the privacy budget for a single timestamp is $\varepsilon_s = \varepsilon_e^t+\varepsilon_c^t+\varepsilon_i^t$,
and the total privacy budget in $w$ timestamps is $\varepsilon = w\cdot\varepsilon_s$. 
We can obtain the following theorem, and the proof is deferred to \autoref{sec:proof_appendix} due to the space limitation.

\begin{theorem}
\label{throrem:PrivGraph}
\method satisfies $w$-event $\varepsilon$-edge DP, where $\varepsilon=w\cdot\varepsilon_s$ and $\varepsilon_s=\varepsilon_e^t + \varepsilon_c^t + \varepsilon_i^t$.
\end{theorem}

\mypara{Complexity Analysis}
We compare the time complexity and the space complexity of various methods.
\hrg has the highest time complexity while \der has the 
highest space complexity. 
The corresponding analysis is in \autoref{sec:complexity_analysis}.

\section{Evaluation}
\label{sec:evaluation}

In~\autoref{subsec:end_to_end}, we first conduct an end-to-end experiment to illustrate the effectiveness of \method compared to existing state-of-the-art methods for static graph publishing.
The baselines are extended to stream scenarios in the same way, which is introduced in~\autoref{section:existing_solutions}.
Then, we perform ablation experiments to explore the influence of each component of~\method. 
Next, we verify the impact of the parameter setting of~\method.
Furthermore, we explore the scalability of different methods.
We also provide the performance of various methods at each timestamp in~\autoref{subsec:appendix_detailed_analysis}.

\subsection{Experimental Setup}
\label{subsec:experimental_setup}

\mypara{Datasets}
We evaluate the performance of different methods on the four real-world datasets. 
\autoref{table:dataset_statistics} provides the total number of nodes and edges from all timestamps of the datasets, 
and the details
are deferred to~\autoref{appendix_datasets}.

\mypara{Metrics}
Here, we evaluate the quality of the synthetic graphs from three key perspectives:  spectral characteristics (eigenvalue nodes), topological structure (assortativity coefficient, degree distribution), and statistical properties (density, clustering coefficient)~\cite{newman2003structure}.
These metrics are also widely adopted by existing DP graph synthesis works~\cite{qin2017ldpgen,nguyen2015differentially,yuan2023privgraph}.
We provide details of the metrics in~\autoref{appendix_metrics}.

\begin{table}[!t]
    \caption{Dataset Statistics.}
    \vspace{-0.4cm}
    \centering
    \footnotesize
    \setlength{\tabcolsep}{0.7em}
    \begin{tabular}{c | c | c | c | c}
    \toprule
     \textbf{Dataset} & \textbf{Nodes} & \textbf{Edges} & \textbf{Timestamps} & \textbf{Type} \\
     \toprule
       As-733~\cite{leskovec2005graphs} & 7,473 & 22,705 & 147 & Communication \\
       As-caida~\cite{leskovec2005graphs} & 31,092 & 97,164 & 25 & Communication \\
        Cit-HepPh~\cite{leskovec2007graph} & 12,905 & 764,525 & 36 & Citation  \\
        Reddit~\cite{kumar2018community} & 34,191 & 125,162 & 24 & Hyperlink \\

      \bottomrule
    \end{tabular}
    \vspace{-0.5cm}
    \label{table:dataset_statistics}
\end{table}

\mypara{Competitors} 
To our knowledge, there are no other differential privacy synthesis approaches optimized for streaming graphs, 
and we adopt existing differential privacy synthesis methods for the static graph as the baselines. 
Specifically, we evenly allocate the total privacy budget across each timestamp. 
Subsequently, we perturb the original graph individually using these differential privacy synthesis methods for the static graph. 
The baselines include \hrg~\cite{xiao2014differentially}, \der~\cite{chen2014correlated},  \tmf~\cite{nguyen2015differentially}, \gen~\cite{qin2017ldpgen}, and \privg~\cite{yuan2023privgraph}. 
In~\autoref{sec:difference_from_privgraph}, we highlight the differences between this work and~\privg.
In addition, we provide a detailed 
description of all baselines
in~\autoref{appendix_baselines}. 
To ensure a fair comparison, we adopt the recommended parameters from the original papers. 

\mypara{Experimental Settings}
For \method, we set the window size $w=5$, and set the threshold parameter for whether to re-partition the communities to $N^t$ (the number of nodes) in the community determination phase.

\mypara{Implementation}
We explore the performance of various methods with a total privacy budget $\varepsilon$ ranging from 0.5 to 4.0. 
All experiments are implemented on a server with AMD EPYC 7402 24-Core processor and 512GB memory.
For each metric, we compute the mean value across all timestamps as a single experimental result. 
Then, we repeat the experiment 10 times for each setting and report the mean and the standard variance. %

\subsection{End-to-End Evaluation}
\label{subsec:end_to_end}

\begin{figure*}[!t]
\centering
\includegraphics[width=0.75\textwidth]
{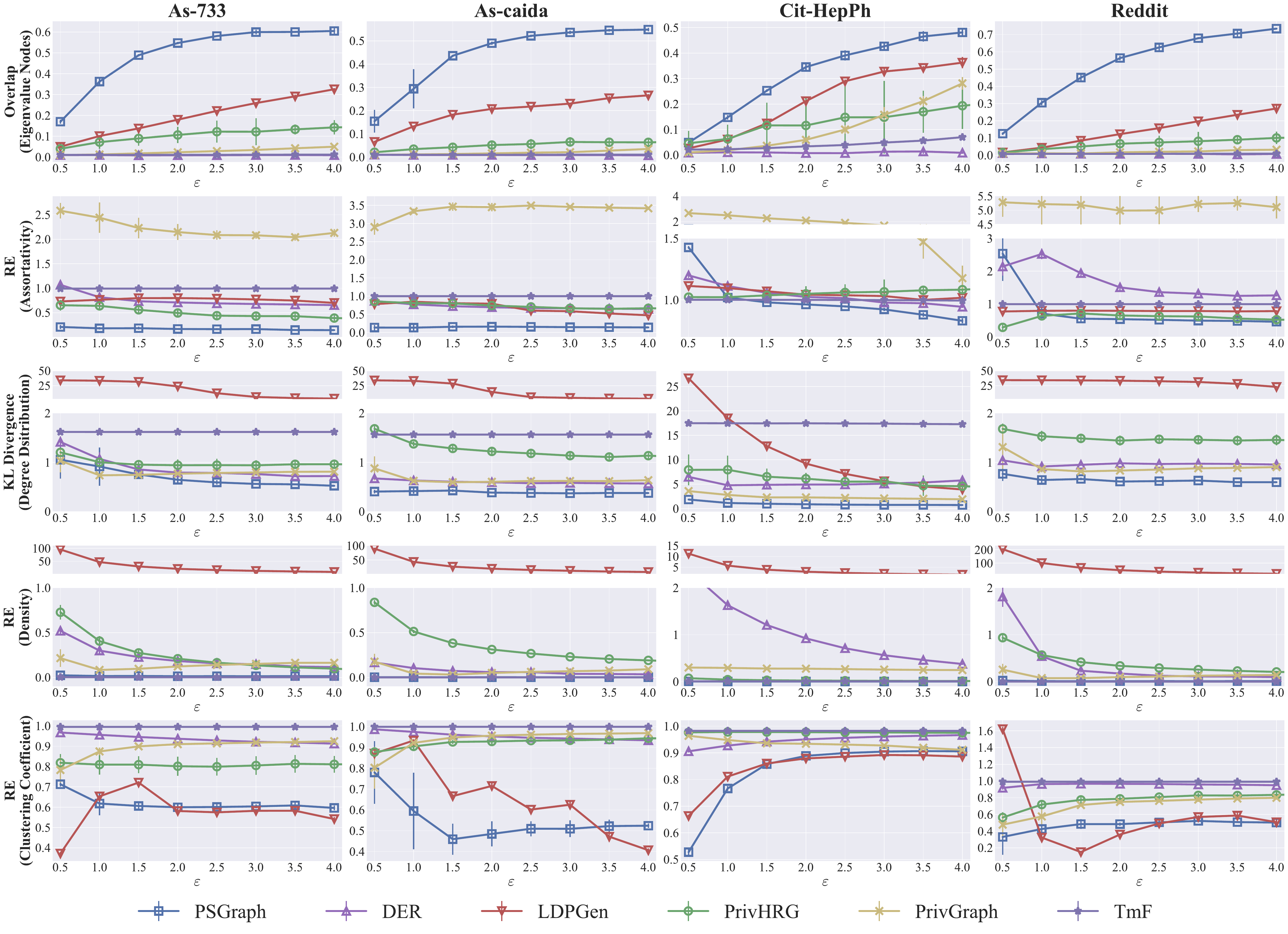}
\vspace{-0.4cm}
\caption{
{End-to-end comparison of different methods.
The columns represent the used datasets, and the rows stand for different metrics.
In each plot, the $x$-axis denotes the privacy budget $\varepsilon$, and the $y$-axis denotes the performance. For the first row, higher is better. For the last four rows, lower is better.}}
\label{fig:end_to_end}
\vspace{-0.4cm}
\end{figure*}

In the section, we perform an end-to-end evaluation of \method and the baselines on five metrics. 
\autoref{fig:end_to_end} illustrates the experimental results on four different datasets.

\mypara{Results on Eigenvalue Nodes}
The first row of \autoref{fig:end_to_end} shows the overlap of nodes in the top 1\% eigenvalues between the original and generated graphs on four datasets, where a higher value stands for higher accuracy. 
We can obtain the following observations from the results.
1) \method outperforms other methods across all datasets. 
By obtaining perturbation degree information both within and outside the community, \method effectively preserves the characteristics of influential nodes. 
Moreover, \method fully leverages streaming data attributes, leading to a more judicious allocation of privacy budget and a higher overlap of feature nodes.
2) \gen and \hrg perform better than other baseline methods because they incorporate the probability of node connections during their design process, which aids in recovering eigenvalue nodes.
3) \privg demonstrates better performance on the Cit-HepPh dataset compared to others. 
This is partly due to the dataset's higher density and more pronounced community clustering, which enhance \privg's effect. 
In addition, the small privacy budget allocated per time step adversely affects the performance of \privg.
4) \tmf and \der exhibit the poorest performance. 
The designs of these methods
involve random edge assignments within specific regions, which are less conducive to preserving eigenvalue nodes.

\mypara{Results on Assortativity Coefficient}
The second row depicts the relative error (RE) of assortativity coefficients between original and synthetic streaming graphs.
We obtain the following key observations.
1) \method outperforms other methods in most cases, especially with higher privacy budgets.
This superiority stems from \method's effective preservation of node features by leveraging the degree information of nodes both within and outside communities and accurately capturing dynamic variations.
For the Cit-HepPh dataset, which is a dense network of edges, lower privacy budgets lead to greater disruption of real edges, resulting in higher errors in assortativity coefficients. 
For the Reddit dataset, the amplitude of this metric is small, resulting in higher relative errors under significant disturbances.
2) \privg shows the weakest performance overall. 
The reason is that \privg only focuses on protecting the connections between nodes within the community, ignoring the connections between communities, which affects the final recovery.
3) \tmf exhibits a RE close to 1 across all four datasets due to its random edge assignment approach, lacking node information protection. 
This randomness leads to higher errors under low privacy budgets.
4) Other mechanisms include \hrg, \gen, and \der do not consistently perform well on all datasets but show efficacy on specific datasets. 
Although these methods consider node connectivity in their designs, their applicability across diverse datasets remains limited.

\mypara{Results on Degree Distribution}
We explore the performance of various methods on the nodes' degree distribution.
We can obtain the following observations from the third row of~\autoref{fig:end_to_end}.
1) \method exhibits significantly lower KL divergence across various datasets compared to other methods. This superiority can be attributed to the extraction of node degree information. 
Furthermore, through information fusion and effective post-processing, \method achieves a high degree of similarity in nodes' degree distribution between the original and generated graphs.
2) \privg also demonstrates relatively low KL divergence across different datasets. 
This is due to \privg's focus on extracting edge information within communities, which aids in preserving the degree distribution of the original graph.
3) \der performs well in degree distribution relative to other baseline methods, except for the Cit-HepPh dataset. 
The denser connections and higher-degree nodes in the Cit-HepPh dataset make it more susceptible to the random edge assignment of \der, potentially disrupting original node characteristics.
4) \hrg and \tmf perform less favorably on this metric compared to most other methods, as their designs do not prioritize nodes' degree preservation.
5) \gen exhibits particularly high KL divergence because significant perturbations under small privacy budgets can lead to a substantial increase in edge counts compared to the original graph. 
This discrepancy significantly impacts the node degree distribution of the generated graph.

\mypara{Results on Density}
The fourth row shows the RE of each method in density, yielding the following observations:
1) \method and \tmf excel in density metric, achieving errors close to zero. 
Both methods perturb the true edge count from the original graph and utilize it as a crucial constraint in their graph synthesis process.
2) \privg also demonstrates relatively small density errors. 
By protecting the edge counts within and between communities, \privg maintains a density close to that of the original graph.
3) \hrg outperforms \der on the Cit-HepPh dataset but performs less effectively on the other three datasets. 
This disparity is primarily due to the dense edge structure of the Cit-HepPh dataset compared to the sparse edges in others. 
\der tends to disrupt the original graph's characteristics more severely under denser edge conditions.
4) \gen shows the poorest performance in density. 
This is because it generates a high number of edges, particularly under low privacy budgets, resulting in a significantly higher density in the synthetic graph compared to the original graph.

\mypara{Results on Clustering Coefficient}
The last row of~\autoref{fig:end_to_end} evaluates the performance of all methods on clustering coefficients, leading to the following observations:
1) \method and \gen demonstrate superior performance compared to other methods. 
\method effectively captures edge information within and between communities, 
employing meticulous post-processing to enhance the fidelity of the synthesized graph, 
thereby achieving better clustering coefficient performance. 
\gen utilizes $k$-means clustering for node grouping, which aids in preserving true clustering coefficients. 
It's noteworthy that a larger privacy budget does not mean that the RE of the clustering coefficients is smaller for all datasets.
This trend is influenced by dataset characteristics. 
The As-caida and Cit-HepPh datasets exhibit higher clustering coefficients, with the latter being denser. 
Consequently, the performance trends of \gen and \method are relatively consistent in these datasets but diverge in others.
The clustering coefficients of the other two datasets are relatively small, so the variations in \gen on these two datasets are more drastic.
2) The overall performance of \privg and \hrg is better than other baselines.
The node's community partitioning of \privg and the node's probabilistic connectivity design of \hrg contribute to the restoration of true clustering coefficients.
3) \der performs only better than \tmf. 
They do not fully consider the connection conditions between different nodes. 
\tmf randomly allocates edges across the entire adjacency matrix, while \der allocates edges within a certain region.

\mypara{Takeaways}
Based on the analysis and results, we can obtain the following conclusions. 
\begin{itemize}[itemsep=0pt,topsep=2pt,parsep=0pt]
    \item \method partitions nodes into communities, perturbs degree information within and between communities, and employs essential post-processing during reconstruction, which effectively preserves assortativity coefficients, density, and clustering coefficients of the real graph.
    \item \method dynamically evaluates variations in the streaming graphs to decide whether to re-partition communities, integrating information across different timestamps when changes are small. 
    This approach significantly enhances the overlap ratio of eigenvalue nodes and reduces the KL divergence in degree distribution.
    \item \gen utilizes $k$-means clustering to partition nodes, achieving good performance in clustering coefficients and eigenvalue nodes.
    \privg focuses on extracting intra-community node edges, thereby achieving a low KL divergence of degree distribution. 
    \tmf reconstructs graphs based on the number of perturbed edges, maintaining a density close to that of the original graph. 
    \der groups similar nodes and randomly assigns edges within specific regions, which performs better in the graphs with sparse edges.
    \hrg considers edge connection probabilities between different nodes, resulting in an outstanding performance in terms of assortativity coefficients. 
\end{itemize}

\subsection{Ablation Study}
\label{subsec:ablation}

\begin{figure*}[!t]
    \centering
    \includegraphics[width=0.75\textwidth]{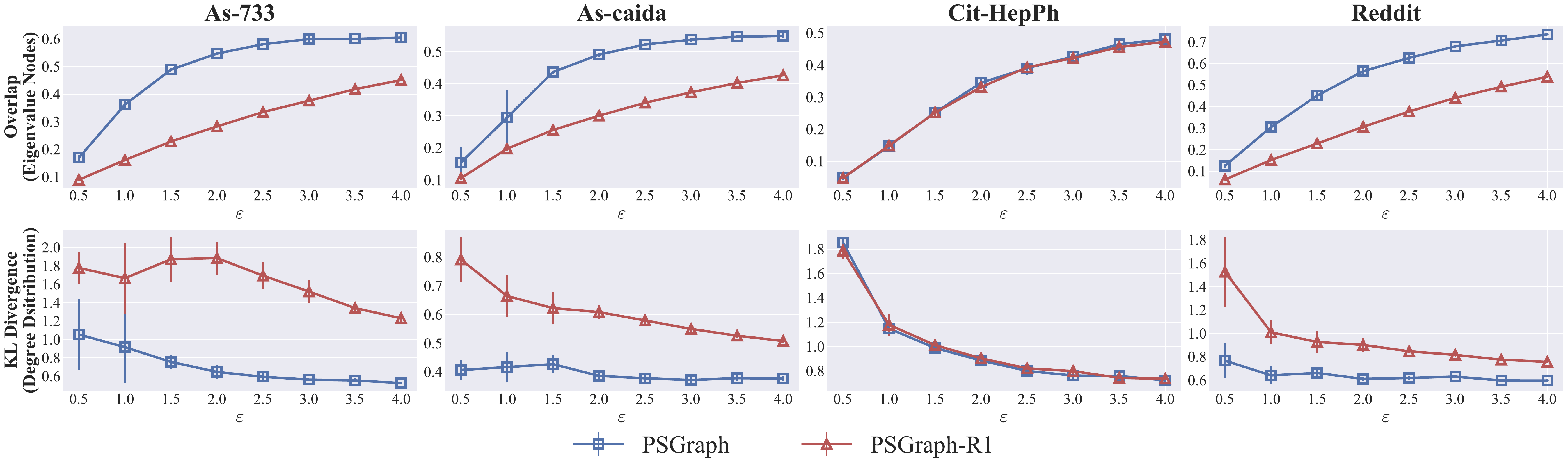}
    \vspace{-0.3cm}
    \caption{
    Comparison of \method and \method-R1. 
    The columns represent the used datasets, and the rows stand for different metrics.
    In each plot, the $x$-axis denotes the privacy budget $\varepsilon$, and the $y$-axis denotes performance.
    For the first row, higher is better. For the last row, lower is better.
    }
    \label{fig:ablation_repart_comm_2_metric}
\vspace{-0.1cm}
\end{figure*}

\mypara{Impact of the Community Judgment}
Recalling~\autoref{subsec:community_determination}, in the community determination phase, 
\method assesses whether there has been a significant change in the current graph compared to the previous timestamp. 
If the change is small, the previously partitioned community remains unchanged; otherwise, the community undergoes re-partitioning. 
In this section, we validate the effectiveness of the community judgment. 
Specifically, during the phase of community determination, we perform re-partitioning each time (called \method-R1) and keep the other procedures unchanged.
Then, we compare the performance between \method and \method-R1.
\autoref{fig:ablation_repart_comm_2_metric} illustrates the 
comparison results on the two metrics.

\method demonstrates significantly superior performance compared to \method-R1 across three datasets, with the exception of the Cit-HepPh dataset. 
The reason is that the Cit-HepPh dataset exhibits low relationships between various timestamps, unlike the other three datasets, which display high coherence. 
Consequently, \method-R1 does not yield a pronounced advantage in the Cit-HepPh dataset.
In contrast, \method effectively leverages the temporal coherence present in the other datasets. 
\method retains previous community partitions in moments of small variations, thus conserving the privacy budget for information perturbation in the subsequent phase. 
This operation ensures the utility of synthetic streaming graphs. 
From this, we find that whether streaming graphs with drastic changes or streaming graphs with low variations,
\method can achieve outstanding performance, which underscores the importance of community judgment.
We also provide the comparison results on the other three metrics in~\autoref{sec:appendix_ablation_study}.

\mypara{Impact of Various Components}
Recalling~\autoref{subsec:information_pert} and~\autoref{subsec:graph_recons},
there are two main components in the second and third phases:
The information fusion step in the information perturbation phase,
and the edge post-processing step in the graph reconstruction phase.
We verify the effectiveness of these two steps with four variants, 
which are summarized in~\autoref{table:detail_ablation_study}. 

\begin{figure*}[!t]
    \centering
    \includegraphics[width=0.75\textwidth]{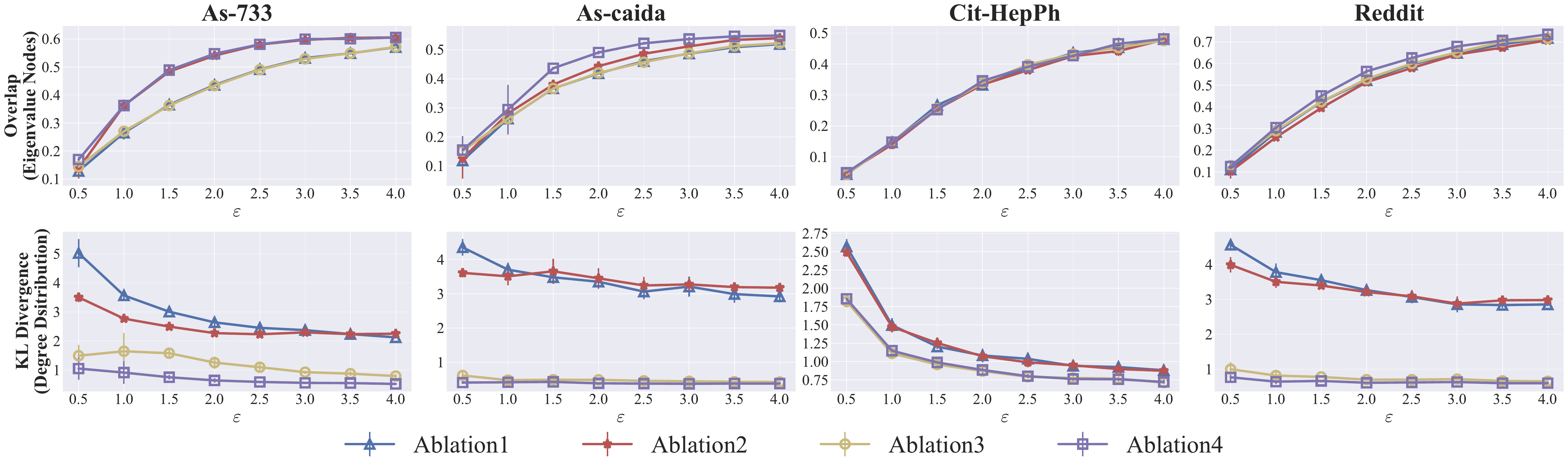}
    \vspace{-0.3cm}
    \caption{
    Effectiveness of various components.
    Ablation1 represents \method without information fusion and post-processing, Ablation2 represents \method without post-processing,
    Ablation3 represents \method without information fusion,
    and Ablation4 represents \method.
    The columns represent the used datasets, and the rows stand for the metrics.
    In each plot, the $x$-axis denotes the privacy budget $\varepsilon$ and the $y$-axis denotes performance.
    For the first row, higher is better. For the last row, lower is better.
    }
    \label{fig:ablation_component_2_metric}
\end{figure*}

\autoref{fig:ablation_component_2_metric} illustrates the performance of various components on two metrics,
and the results on the other three metrics are deferred to~\autoref{sec:appendix_ablation_study}.
From~\autoref{fig:ablation_component_2_metric},
Ablation1, excluding both information fusion and post-processing, shows inferior results on both metrics.
Ablation2 and Ablation3, incorporating either information fusion or post-processing alone, exhibit a certain improvement compared to Ablation1.
Ablation4 achieves the best performance by integrating both information fusion and post-processing, highlighting the pivotal role of these steps.
Moreover, due to the drastic changes in the Cit-HepPh dataset, the impact of information fusion is less pronounced on this dataset.

\begin{table}[!t]
    \caption{Details of ablation studies.}
    \vspace{-0.4cm}
    \centering
    \footnotesize
    \begin{tabular}{c | c | c }
    \toprule
     \textbf{Name} & \textbf{Information fusion} & \textbf{Post-processing}  \\
     \toprule
       Ablation1 & $\times$ & $\times$ \\
       Ablation2 & $\checkmark$ & $\times$ \\
       Ablation3 & $\times$ & $\checkmark$ \\
       Ablation4 & $\checkmark$ & $\checkmark$ \\
      \bottomrule
    \end{tabular}
    \vspace{-0.3cm}
    \label{table:detail_ablation_study}
\end{table}

\subsection{Parameter Variation}
\label{subsec:parameter_variation}

\begin{figure*}[!t]
    \centering
    \vspace{-0.2cm}
    \includegraphics[width=0.75\textwidth]{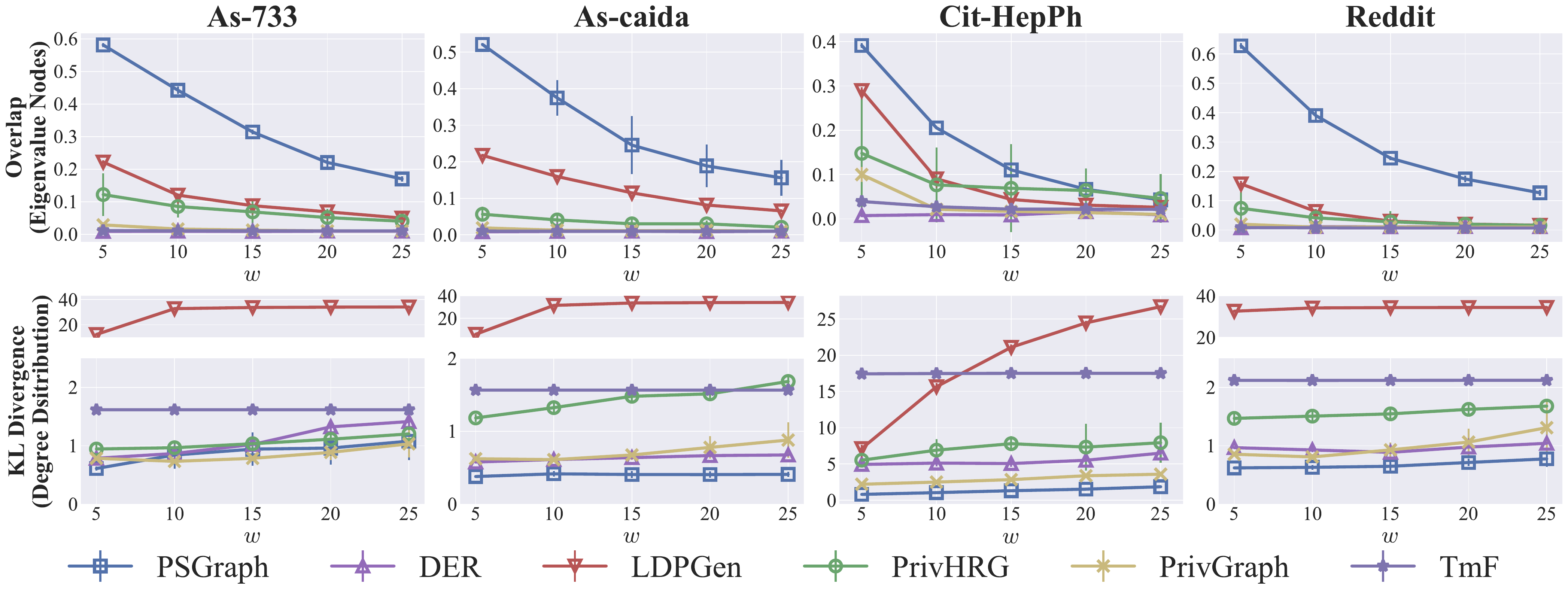}
    \vspace{-0.4cm}
    \caption{
    Impact of window size. 
    The columns represent the used datasets, and the rows stand for different metrics.
    In each plot, the $x$-axis denotes the number of window size $w$ and the $y$-axis denotes performance.
    For the first row, higher is better. For the last row, lower is better.
    }
    \label{fig:vary_w_2_metric}
    \vspace{-0.2cm}
\end{figure*}

\mypara{Impact of Window Size}
\autoref{fig:vary_w_2_metric} illustrates the performance of different methods across various window size $w$ with a privacy budget of 2.5. 
Due to space constraints, we provide a comparison of the overlap ratio of eigenvalue nodes and the KL divergence of degree distribution. 
The results highlight the significant advantages of \method over other methods, showcasing the effectiveness of our dynamic community determination mechanism and post-processing steps in allocating privacy budget and enhancing utility.
Additionally, as $w$ increases, the overlap of eigenvalue nodes exhibits a declining trend, mainly due to the reduced privacy budget allocated to each timestamp, posing a challenge to graph reconstruction. 
Nonetheless, \method still maintains great performance compared to other baselines.
We also provide the performance of other metrics at various $w$ in~\autoref{subsec:appendix_parameter_variation}.

\begin{figure*}[!t]
    \centering
    \includegraphics[width=0.75\textwidth]{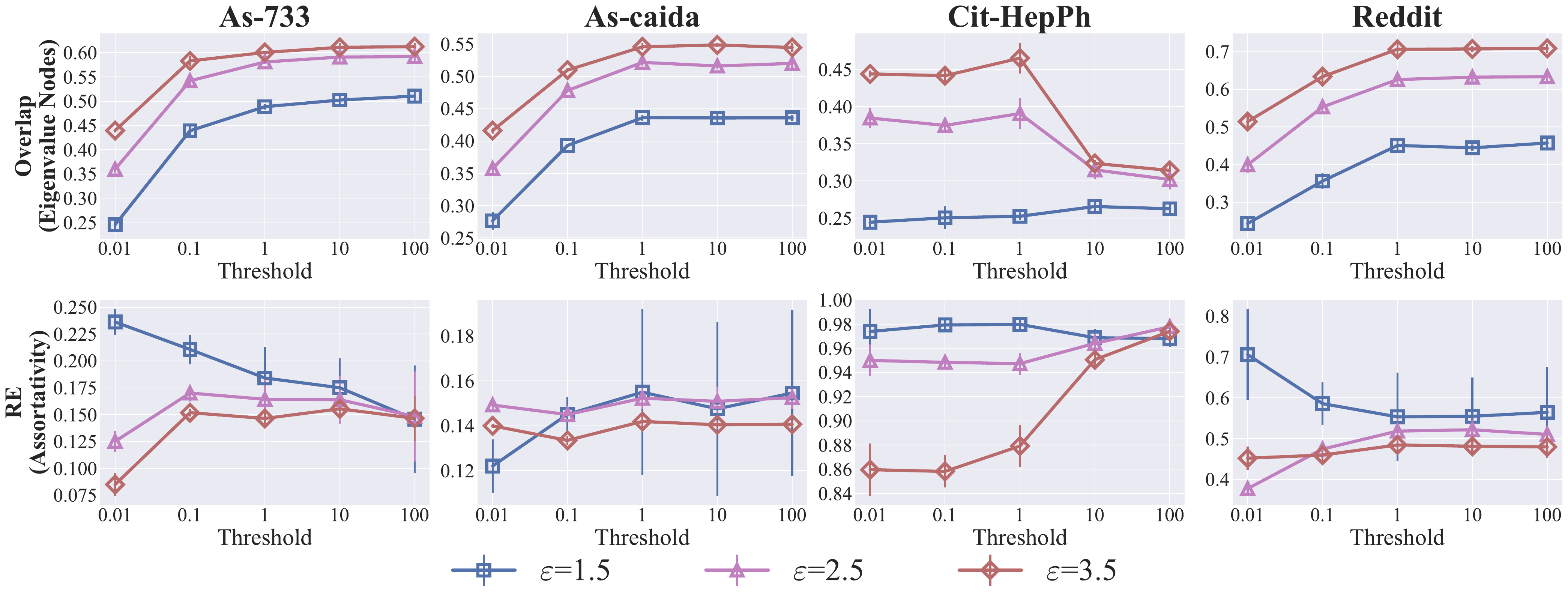}
    \vspace{-0.5cm}
    \caption{
    Impact of threshold. 
    The columns represent the used datasets, and the rows stand for different metrics.
    In each plot, the $x$-axis denotes the threshold and the $y$-axis denotes performance.
    For the first row, higher is better. For the last row, lower is better.
    }
    \label{fig:vary_threshold_2_metric}
    \vspace{-0.4cm}
\end{figure*}

\mypara{Impact of Threshold}
Recalling~\autoref{subsec:community_determination}, 
\method determines whether to re-partition communities based on variations in edge counts between current and previous timestamps. 
In this section,
our aim is to explore the impact of different threshold settings on various metrics.
\autoref{fig:vary_threshold_2_metric} illustrates the REs of eigenvalue node overlap and assortativity coefficients across different privacy budgets, with thresholds ranging from 0.01 times to 100 times the number of nodes. 
Due to the space constraints, the performance of the other three metrics under various thresholds is provided in~\autoref{subsec:appendix_parameter_variation}. 
Our observations are as follows.
1) Datasets show inconsistent performance under the same threshold. 
For instance, As-733, As-caida and Reddit datasets exhibit lower eigenvalue node overlap ratios with smaller thresholds, whereas the Cit-HepPh dataset shows higher overlap ratios under similar conditions. 
This divergence stems from Cit-HepPh's more pronounced changes and lower relationships between adjacent timestamps compared to the other datasets.
2) The same metric on a single dataset demonstrates varying trends with different privacy budgets. 
For example, when $\varepsilon=1.5$, the RE of assortativity coefficient in the As-733 dataset decreases steadily with increasing threshold, while when $\varepsilon=3.5$, the RE shows an increasing trend with higher thresholds.
3) Under the same dataset and privacy budget, different metrics exhibit distinct trends. 
For instance, with a privacy budget of 3.5 on the As-733 dataset, small thresholds yield low eigenvalue node overlap ratios and small REs in assortativity coefficient.

Based on these results, we find no single threshold that satisfies all privacy budgets, metrics, and datasets. 
Overall, \method consistently maintains competitive quality when using the number of nodes as the threshold. 
Therefore, this criterion is applied uniformly throughout our experiment.

\subsection{Scalability}
\label{subsec:scalability}
\begin{figure}[!t]
    \centering
    \includegraphics[width=0.40\textwidth]{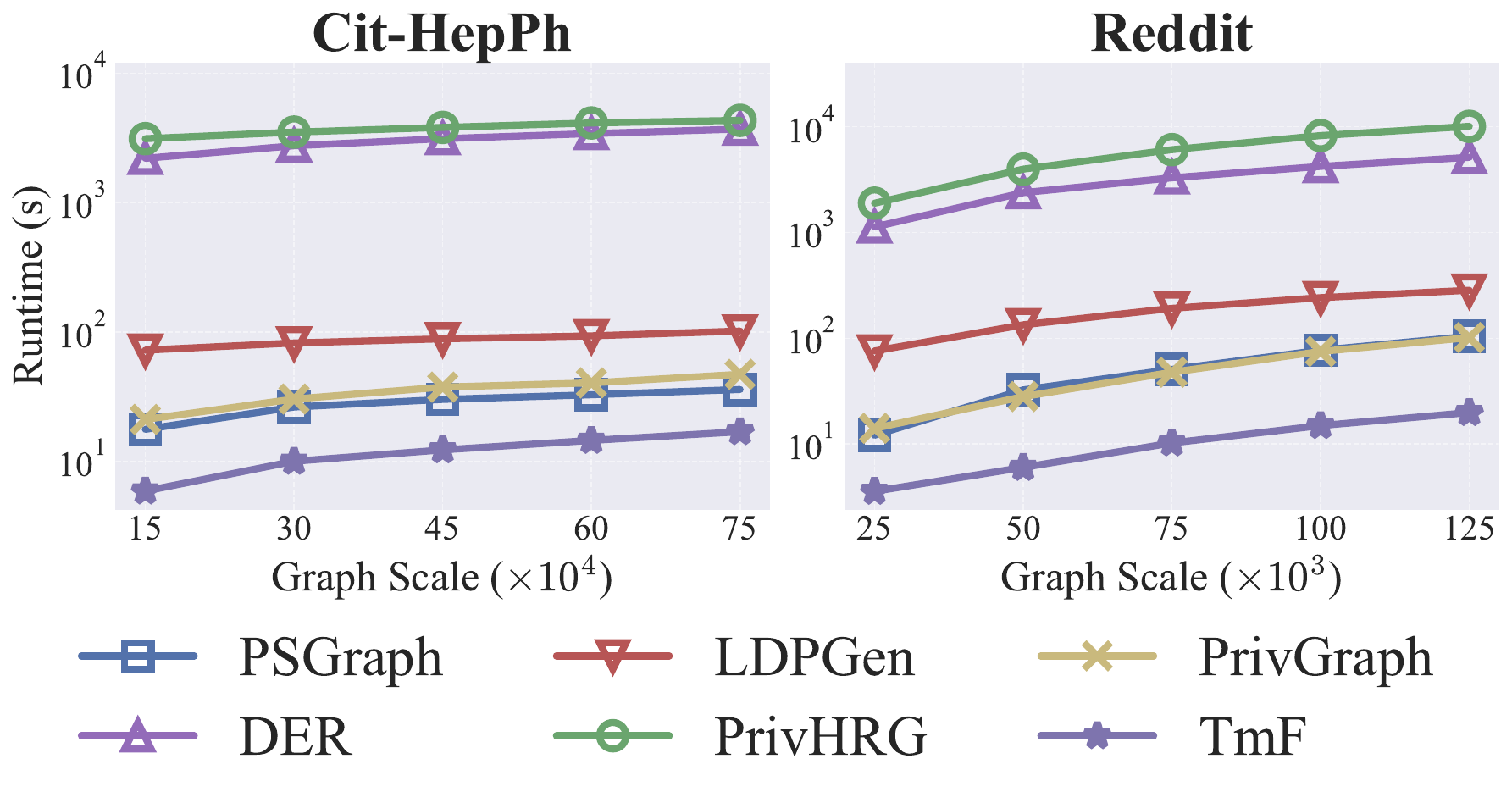}
    \vspace{-0.45cm}
    \caption{
    Scalability evaluation.
    }
    \label{fig:scalability}
    \vspace{-0.6cm}
\end{figure}

We further analyze the scalability by varying the number of edges for all timestamps.
The running time results on the Cit-HepPh and Reddit datasets are illustrated in~\autoref{fig:scalability}.
The runntime of~\tmf is the shortest, but its performance is poor.
\method performs better than most methods,
and its processing time will not grow sharply with the increase of edge scales.
Therefore, \method is suitable for practical deployment with low computational cost.

\section{Related Work}
\label{sec:related_work}

\subsection{Differentially Private Stream Release}
DP
has been widely employed in data streaming scenarios where data is continuously published.
The privacy notion in stream release mainly includes event-level, user-level, and $w$-event privacy. 

\mypara{Event-level Privacy} 
It is utilized to protect individual timestamps~\cite{cao2017quantifying,chen2017pegasus,dwork2010differential,perrier2019private,wang2021continuous}. 
Chen \etal~\cite{chen2017pegasus} use a perturb-group-smooth architecture to answer multiple stream queries. 
Perrier \etal~\cite{perrier2019private} observe that stream data often concentrates below a threshold significantly lower than the upper bound, devising a method based on the smooth sensitivity to identify this threshold. 
Wang \etal~\cite{wang2021continuous} tackle the issue of unbounded maximum values in real-value data stream publishing using a threshold estimation mechanism.

\mypara{User-level Privacy}
Compared to event-level DP, which protects against changes in individual events, user-level privacy aims to hide all events of any user~\cite{acs2014case,dong2023continual,erlingsson2019amplification,fan2013adaptive,li2015differentially}.
Fan \etal~\cite{fan2013adaptive} propose to release perturbed statistics at sampled timestamps and utilize the Kalman filter to predict non-sampled values, thereby correcting the noisy sampled values. 
Another direction involves the offline setting, where the server initially has a global view of all values and subsequently releases a streaming model that satisfies DP. 
In this context, Acs~\etal~\cite{acs2014case} propose an algorithm for releasing spatio-temporal density using the discrete fourier transform.

\mypara{$w$-event Privacy}
To achieve a balance between privacy and utility at both event and user levels, $w$-event privacy is proposed for infinite streams, which protects every running window of at most $w$ timestamps~\cite{bolot2013private,kellaris2014differentially,ma2019real,wang2016real,wang2018privacy}.
Bolot \etal~\cite{bolot2013private} extend the binary tree mechanism for publishing perturbed answers on sliding window sum queries over infinite binary streams with fixed window sizes, introducing a relaxed privacy concept termed decayed privacy.
Kellaris \etal~\cite{kellaris2014differentially} propose the notion of $w$-event DP and two novel privacy budget allocation schemes, where $\varepsilon$ is divided among individual events to achieve the desired privacy level.
Wang \etal~\cite{wang2016real} focus on the release of spatio-temporal trajectories and propose a framework called RescueDP, which incorporates a grouping strategy to partition the dimensions with similar statistics and changing trends to reduce the noise injection.

\subsection{Differentially Private Graph Analysis}

Differentially private graph analysis can perform a variety of statistical tasks on sensitive data. 
There are several works dedicated to specific downstream tasks~\cite{hay2009accurate,raskhodnikova2016lipschitz,upadhyay2021differentially,wang2013spectral,wang2013learning,ye2020towards}, including degree distribution, clustering coefficient, spectral graph analysis, \etc
In particular, lipschitz extensions and exponential mechanism are employed to approximate the degree distribution for a sensitive graph in~\cite{raskhodnikova2016lipschitz}.
A constrained inference method is designed in~\cite{hay2009accurate} to publish the degree distribution. 
In~\cite{wang2013learning}, a divide-and-conquer approach is developed to obtain the clustering coefficient under DP. 
In addition,
Wang \etal~\cite{wang2013spectral} propose two methods to calculate the eigen decomposition of the adjacency matrix while satisfying DP.

\subsection{Differentially Private Data Synthesis}

\mypara{Tabular Data}
There are three mainstream approaches for processing tabular data: graphical model-based, game-based, and deep generative model-based.
The graphical model-based approaches focus on estimating a graphical model which approximates the distribution of the original dataset under DP~\cite{bindschaedler2017plausible,mckenna2019graphical,zhang2017privbayes,zhang2021privsyn,mckenna2022aim,cai2021data}. 
The game-based approaches regard the dataset synthesis problem as a zero-sum game~\cite{gaboardi2014dual,vietri2020new}. 
Assuming that there are two players, \ie, the data player and the query player, 
a no-regret learning algorithm for the data player is proposed to solve the game in \cite{hardt2012simple}. 
Dual Query~\cite{gaboardi2014dual} switches the role of the data player and the query player. 
Deep generative model-based approaches initially train a deep generative model under DP and utilize the model to synthesize the dataset~\cite{beaulieu2019privacy,frigerio2019differentially,zhang2018differentially}. 

\mypara{Trajectory Data}
There are some works that investigate the synthesis of trajectory dataset under DP~\cite{chen2012n-grams,gursoy2018utility,he2015dpt,sun2023synthesizing,wang2023privtrace}. 
He \etal~\cite{he2015dpt} design DPT, which takes into account the non-uniformity of true trajectories. 
AdaTrace 
proposed in~\cite{gursoy2018utility} enforces deterministic constraints on the generated trajectories while satisfying DP, thereby enhancing the algorithm's resilience against common threats.
A method involving dynamic selection of first-order and second-order models is introduced in~\cite{wang2023privtrace} to balance noise and correlation errors.

\section{Conclusion}
\label{sec:conclusion}

In this paper, we introduce \method for publishing streaming graphs under Differential Privacy (DP). 
Our approach effectively preserves streaming graph information by extracting node details at the community level and applying essential post-processing techniques. 
Additionally, by leveraging temporal dynamics in the streaming data and aggregating relevant information across various timestamps, we efficiently conserve the privacy budget, resulting in synthetic graphs with superior utility compared to state-of-the-art methods. 
Extensive experiments on four real-world datasets illustrate the superiority of \method.
We also conduct comprehensive ablation studies and parameter experiments to analyze the impact of various parts of \method.
The results indicate that the close integration of various modules in \method can effectively capture the dynamic variations of the streaming graphs.

\bibliographystyle{ACM-Reference-Format}
\bibliography{easy}

\appendix

\section{Difference from~\privg}
\label{sec:difference_from_privgraph}
We would like to emphasize that our proposed \method is fundamentally different from \privg~\cite{yuan2023privgraph} in the following several key aspects. 
First, the approach to community determination is different. 
\privg is designed for the static graph, requiring communities to be re-identified at each timestamp. 
In contrast, \method dynamically assesses whether community re-division is necessary based on the variation between the current graph and last graph. 
Second, the strategies for node feature extraction and perturbation also diverge. \privg extracts only the degree information of nodes within each community, whereas \method incorporates the node degree information within and outside the community, thereby reducing information loss. 
Third, the graph reconstruction mechanisms are distinct. 
\privg reconstructs inter-community edges using random connections, while \method leverages the degree information of nodes outside the community and the inter-community edge patterns (as shown in~\autoref{eq:prob_inter_community}) to perform a more informed reconstruction. Furthermore, \method applies an edge post-processing step to enhance the fidelity of the synthetic graph.
Therefore, the method design of \method
is essentially different from~\privg.

\section{Proof of Throrem 1}
\label{sec:proof_appendix}

\method consists of three phases: Community determination, information perturbation, and graph reconstruction.
Next,
we show that \method satisfies $w$-event $\varepsilon$-edge DP, where $\varepsilon=w\cdot \varepsilon_s$ and $\varepsilon_s=\varepsilon_e^t+\varepsilon_c^t+\varepsilon_i^t$.

\mypara{Proof 1: Community determination satisfies $(\varepsilon_e^t+\varepsilon_c^t)$-edge DP}
\begin{proof}
    \label{proof:community_determination}
    In the process of community determination, \method first perturbs the true number of edge based on Laplace mechanism, which can provide rigorous differential privacy guarantee. 
    The privacy budget consumed in this step is $\varepsilon_e^t$.    
    Then, the difference in the number of perturbed edges between the current timestamp and the last timestamp determines whether to retain the previous community or re-partition.
    If the communities from last timestamp is retained,
    it does not consume any privacy budget, \ie, $\varepsilon_c^t=0$.
    Conversely, if the community needs to be re-partition,
    the Laplace mechanism is adopted to perturb the weights of supernodes,
    while exponential mechanism is utilized to determine the nodes' final community, both of which meet the requirements of DP.
    The detailed proof of the re-partition can be found in Appendix A of~\cite{yuan2023privgraph}.
    Therefore,
    the community division step satisfies $\varepsilon_c^t$-edge DP.
    Further, the community determination satisfies $(\varepsilon_e^t+\varepsilon_c^t)$-edge DP based on sequential composition of DP.
    
\end{proof}

\mypara{Proof 2: Information perturbation satisfies $\varepsilon_i^t$-edge DP}
\begin{proof}
    \label{proof:information_perturbation}
    In the information perturbation phase, the degree information within and outside the community, and the edge count between various communities are injected Laplace noise, respectively.
    Note that both the degree information outside the community and the edge count between communities require access to the edges of inter-community,
    thus their privacy budget allocation needs to follow sequential property of DP.
    On the other hand, the degree information within the community needs to visit the edges of intra-community, which does not intersect with the above two. 
    Then,
    the degree information of intra-community can be perturbed by the same privacy budget based on the parallel composition.
    Therefore, the information perturbation satisfies $\varepsilon_i^t$-edge DP.
\end{proof}
\mypara{Overall Privacy Budget}
According to the above proofs,
in the first phase,
community determination satisfies $(\varepsilon_e^t+\varepsilon_c^t)$-edge DP. 
The information perturbation phase satisfies $\varepsilon_i^t$-edge DP.
In graph reconstruction, \method processes the perturbed data without consuming privacy budget.
In accordance with sequential composition, \method satisfies $\varepsilon_s$-edge DP at a single timestamp.
Since the privacy budget at each timestamp is the same,
\method satisfies $w$-event $\varepsilon$-edge DP for any $w$ timestamps.

\section{Complexity Analysis}
\label{sec:complexity_analysis}

In the section, we analyze the computational complexity of \method, and quantitatively compare its running time and memory consumption with other methods. 

\mypara{Computational Complexity}
We adopt $m$ to represent the number of edges and $n$ to represent the number of nodes.
For the time complexity of \method, in the first phase, if there are significant variations between adjacent two graphs, it is essential to re-partition the community, which approximately requires the complexity of $\mathcal{O}(m)$.
In the information perturbation phase, the degree information of nodes and the edge count between various communities are perturbed,
thus the corresponding time complexity is $\mathcal{O}(n+k^2)$, where $k$ is the number of communities.
The time complexity of the graph reconstruction phase is similar to that of the second phase.
Above all, we can obtain $\mathcal{O}(m)+\mathcal{O}(n)+\mathcal{O}(k^2)=\mathcal{O}(m+n+k^2)<\mathcal{O}(n^2)$,
thus the total time complexity is $\mathcal{O}(n^2)$.
Actually, $k$ is usually small in practice, and community re-partitioning is not necessary at each timestamp, so the overall runtime is acceptable.
For the space complexity of \method, it requires to store the edge information from all nodes to various communities,
we can obtain $\mathcal{O}(nk)<\mathcal{O}(n^2)$, where $k$ is the number of communities.
Therefore, the space complexity is $\mathcal{O}(n^2)$.
In addition, the time and space complexity of other methods has already been analyzed in~\cite{yuan2023privgraph}, and we omit it here.

\mypara{Empirical Evaluation}
\autoref{table:comparsion_running_time_new} and \autoref{table:comparsion_memory_consumption_new} show the
running time and the memory consumption for all methods on the four datasets (see their details in \autoref{table:dataset_statistics}).
The running time in \autoref{table:comparsion_running_time_new}
illustrates that the performance of \tmf is the best because it perturbs the cells in a linear time.
The running time of \method, \privg and \gen are longer than \tmf due to community division.
However, community partitioning is fast, and it's not always necessary for \method. 
Therefore, \method still performs better compared to most methods.
\hrg and \der take much more time than other methods.
\hrg spends a lot of time to sample an HRG, and \der consumes abundant time dividing the adjacency matrix into small pieces at each timestamp.

\autoref{table:comparsion_memory_consumption_new} illustrates the memory consumption.
We can see that \der occupies the most memory.
This is because \der requires maintaining a count matrix during data processing.
\method also necessitates a substantial amount of memory to store information related to nodes and communities. 
Overall, this level of resource usage is acceptable.

\begin{table}[!t]
\caption{Comparison of running time (measured by seconds).}
    \vspace{-0.3cm}
    \centering
    \footnotesize
    \setlength{\tabcolsep}{0.5em}
    \begin{tabular}{c | c | c | c | c }
    \toprule
    & \multicolumn{4}{c}{\textbf{Datasets}} \\
    {\textbf{Methods}} & {As-733} & {As-caida} & {Cit-HepPh} & {Reddit} \\
     \toprule
       \tmf  & 70.84 & 481.49  & 16.88 & 19.88  \\
       \privg  & 310.89 & 1998.91  & 46.93 & 101.60  \\
       \der & 8772.87 & 48223.78 & 3705.31 & 5140.13  \\
       \hrg & 35281.87 & 270312.32 & 4328.67 & 10091.21 \\
       \gen & 512.87 & 1256.19 & 101.42 & 284.64 \\
       \method & 678.94 & 1242.34 & 35.81 & 104.58 \\
      \bottomrule
    \end{tabular}
    
    \label{table:comparsion_running_time_new}
\end{table}

\begin{table}[!t]
\caption{Comparison of memory consumption (measured by Megabytes).}
    \vspace{-0.3cm}
    \centering
    \footnotesize
    \setlength{\tabcolsep}{0.5em}
    \begin{tabular}{c | c | c | c | c }
    \toprule
    & \multicolumn{4}{c}{\textbf{Datasets}} \\
    {\textbf{Methods}} & {As-733} & {As-caida} & {Cit-HepPh} & {Reddit} \\
     \toprule
       \tmf  & 262.26 & 580.06  & 127.64 & 311.56  \\
       \privg  & 267.47 & 610.24  & 159.23 & 361.23  \\
       \der & 505.12 & 7340.03 & 382.29 & 1120.26 \\
       \hrg & 282.15 & 694.23 & 173.21 & 372.19 \\
       \gen & 318.95 & 1604.09 & 183.07 & 455.86 \\
       \method & 385.09 & 2367.49 & 295.12 & 767.32 \\
      \bottomrule
    \end{tabular}
    
    \label{table:comparsion_memory_consumption_new}
    \vspace{-0.2cm}
\end{table}

\section{Experimental Setup}

\subsection{Datasets}
\label{appendix_datasets}
The details of four datasets are as follows.

\begin{itemize}
[itemsep=2pt,topsep=2pt,parsep=0pt]
\item \textbf{As-733~\cite{leskovec2005graphs}.} 
The graph of routers that form the Internet can be structured into sub-graphs known as Autonomous Systems (AS). 
Each AS interacts with neighboring ASes by exchanging traffic flows.
From the BGP (Border Gateway Protocol) logs,
a communication network of who-talks-to-whom can be constructed.
The dataset contains 733 daily instances from November 8, 1997 to January 2, 2000.
Due to the large total number of timestamps, we selected 147 timestamps by taking them every 5 intervals.
The processed dataset contains a total of 7,473 nodes (\ie, routers) and 22,705 edges (\ie, communications).
\item \textbf{As-caida~\cite{leskovec2005graphs}.}
This dataset contains 122 CAIDA AS graphs, from January 2004 to November 2007.
We select 25 of them as the test dataset at intervals of each 5 time points.
The final dataset includes a total of 31,092 nodes (\ie, routers) and 97,164 edges (\ie, communications).
\item \textbf{Cit-HepPh~\cite{leskovec2007graph}.}
Cit-HepPh is from the e-print arXiv and covers all the citations.
We obtain a citation graph for a total of 36 months, from January 1996 to December 1999, based on monthly statistics.
The full citation graph contains 12,905 nodes (\ie, papers) and 764,525 edges (\ie, citations).
\item \textbf{Reddit~\cite{kumar2018community}.}
The dataset is extracted from the posts that create hyperlinks from one subreddit to another.
We choose the hyperlink network by month for a total of 24 months from January 2014 to December 2015.
The overall reddit dataset includes 34,191 nodes (\ie, posts) and 125,162 edges (\ie, hyperlinks).

\end{itemize}

\subsection{Evaluation Metrics}
\label{appendix_metrics}

We evaluate the quality of the generated graph based on the following five different metrics. 

\begin{itemize}
[itemsep=2pt,topsep=2pt,parsep=0pt]

    \item \textbf{Eigenvalue Nodes.}
    The eigenvector centrality score (which can be computed by the principal eigenvector of the adjacency matrix) is utilized to rank the nodes, identifying the most influential nodes within a graph. 
    This metric reflects the similarity of the distribution of key nodes.
    Here, we compare the percentage of common nodes in the top 1\% most influential nodes of the original graph and synthetic graphs. 
    \begin{equation*}
        {Overlap}_{Node} = \frac{\left | N_1 \cap \hat{N}_1 \right |}{\left | N_1 \right |},
    \end{equation*}
    where $N_1$ and $\hat{N}_1$ represent the node sets of the top 1\% eigenvalues in the original graph and synthetic graphs. 

    \item \textbf{Assortativity Coefficient.}
    The metric~\cite{gong2012evolution} reflects the preference pattern of node connections.
    We calculate the relative error (RE) between the original and generated graphs.
    \begin{equation*}
        {RE}_{Ass} = \frac{\left | \hat{A} - A  \right |}{\max(\gamma,A)}, 
    \end{equation*}
    where ${A}$ and $\hat{A}$ are the assortativity coefficients of the original graph and the generated graph, respectively, 
    and $\gamma$ is a small constant to avoid a zero denominator. 
    \item \textbf{Degree Distribution.}
    The node degree plays a critical and intuitive role in measuring nodes' influence and evolution patterns~\cite{costa2007characterization}.
    We adopt the Kullback-Leibler (KL) divergence~\cite{kullback1997information} to measure the difference of the degree distributions between the original graph and the generated graph. 
    \begin{equation*}
        D_{KL}\left(Q \parallel \hat{Q}\right) = \sum_{x\in \mathcal{X}}Q(x)\log
        \left(\frac{Q(x)}{\hat{Q}(x)}\right),
    \end{equation*}
    where $Q(x)$ and $\hat{Q}(x)$ represent the degree distributions of the original graph and the generated graph separately. 

    \item \textbf{Density.}
    Density can reflect the global connectivity of the graph.
    Here, we provide the RE of the density according to the number of nodes and edges between the original and generated graphs.
    \begin{equation*}
        {RE}_{Den} = \frac{\left | \hat{D}_1 - D_1  \right |}{\max(\gamma,D_1)}, 
    \end{equation*}
    where ${D_1}$ and $\hat{D}_1$ are the densities of the original graph and generated graphs, respectively, 
    and $\gamma$ is a small constant to avoid a zero denominator.  

    \item \textbf{Clustering Coefficient.}
    The clustering coefficient can measure the connectivity between neighboring nodes.
    The RE of clustering coefficient can be calculated as follows.
    \begin{equation*}
        {RE}_{CC} = \frac{\left | \hat{Y} - Y  \right |}{\max(\gamma,Y)}, 
    \end{equation*}
    where ${Y}$ and $\hat{Y}$ are the clustering coefficients of the original graph and synthetic graphs, respectively, 
    and $\gamma$ is a small constant to avoid a zero denominator.

\end{itemize}

\subsection{Baselines}
\label{appendix_baselines}

The details of different baseline methods are as follows.

\begin{itemize}
[itemsep=2pt,topsep=2pt,parsep=0pt]

\item \mypara{TmF~\cite{nguyen2015differentially}}
\tmf initially perturbs the adjacency matrix of the original graph by injecting Laplace noise to every cell. 
Subsequently, it selects the top-$m$ noisy cells as edges from the perturbed adjacency matrix, 
where $m$ is determined by injecting Laplace noise to the actual number of edges to satisfy edge-DP. 
However, perturbing all cells in the adjacency matrix introduces excessive noise, leading to the loss of many true edges among the top-$m$ noisy cells, particularly when $\varepsilon$ is small.

\item \mypara{DER~\cite{chen2014correlated}} 
\der primarily involves three main steps: Node relabeling, dense region exploration, and edge reconstruction.
Firstly, \der clusters the edges within the adjacency matrix during the node relabeling step.
Next, \der divides the adjacency matrix into multiple blocks and estimates their density using a noisy quadtree.
Finally, \der reconstructs the edges within every block based on exponential mechanism.

\item \mypara{PrivHRG~\cite{xiao2014differentially}}
\hrg utilizes a statistical Hierarchical Random Graph (HRG)~\cite{clauset2008hierarchical} to model the graph structure, aimed at reducing noise impact. 
The likelihood of an HRG for a graph $G$ indicates the degree of plausibility that the HRG accurately represents $G$.
The method begins by mapping all nodes into a hierarchical structure and capturing connection probabilities between any pair of nodes in the graph. 
\hrg then employs the Markov Chain Monte Carlo method to obtain an HRG that maximizes likelihood while adhering to edge-DP. 
Finally, edges are reconstructed on the basis of perturbed probabilities.

\item \mypara{PrivGraph~\cite{yuan2023privgraph}}
\privg initially constructs a hyper-node weight graph by randomly partitioning nodes. 
Subsequently, it applies community detection algorithm on this perturbed weight graph and adjusts the final community partition using exponential mechanism. 
Following this, \privg extracts the degree sequence of nodes within the community and the number of edges connected between communities, and perturbs them using the Laplace mechanism.
In the end, \privg reconstructs the edges within and between communities based on different edge probability formulas.

\item \mypara{LDPGen~\cite{qin2017ldpgen}}
\gen is originally developed to generate a synthetic graph under local differential privacy (LDP). 
It begins by randomly dividing all nodes into two groups. 
Next, \gen adjusts the partition of all nodes using $k$-means clustering based on the number of connected edges from the nodes to each group. 
Finally, it reconstructs the connected edges based on the grouping and connection information.

\end{itemize}

\section{Ablation Study}
\label{sec:appendix_ablation_study}

\begin{figure*}[!t]
    \centering
    \includegraphics[width=0.75\textwidth]{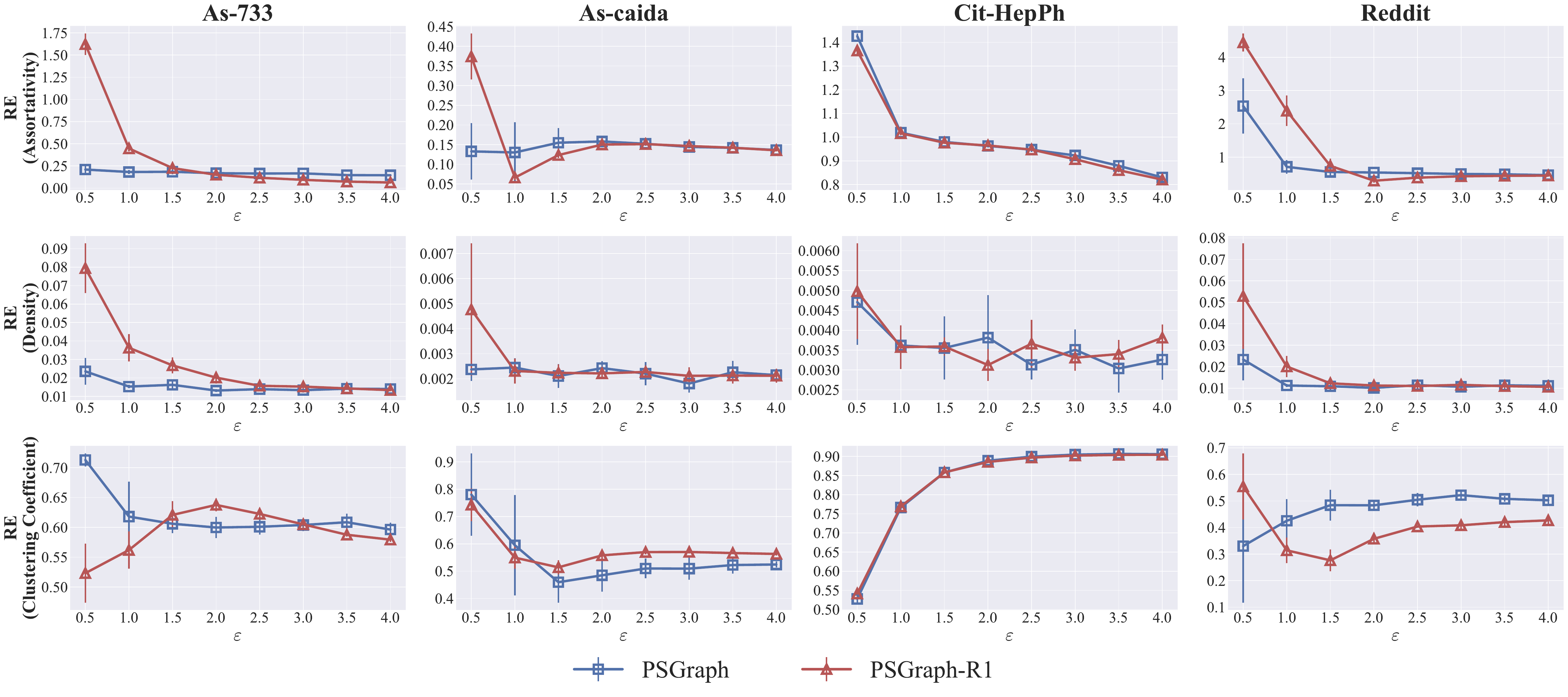}
    \vspace{-0.4cm}
    \caption{
    [Lower is better]
    Comparison of \method and \method-R1. 
    The columns represent the used datasets, and the rows stand for different metrics.
    In each plot, the $x$-axis denotes the privacy budget $\varepsilon$ and the $y$-axis denotes performance.
    }
    \vspace{-0.2cm}
    \label{fig:ablation_repart_comm_3_metric}
\end{figure*}

\mypara{Impact of the Community Judgment}
\autoref{fig:ablation_repart_comm_3_metric} illustrates the comparsion result of \method and \method-R1 across three metrics. 
Regarding the assortativity coefficient, \method maintains consistently low errors on the As-733 and As-caida datasets. In contrast, \method-R1 exhibits significant errors when the privacy budget is limited, due to the substantial privacy budget required for each community re-partitioning. 
This disparity is particularly pronounced under low privacy budget conditions. 
For the Cit-HepPh dataset, where temporal dynamics are less clear, \method and \method-R1 perform similarly, reflecting the dataset's characteristics.
\method also demonstrates advantages in density preservation, underscoring the significance of the community judgment mechanism. 
While \method-R1 occasionally shows improved performance in clustering coefficients, attributed to the beneficial effects of periodic community re-partitioning on node clustering, \method remains competitive performance.
These results highlight the robust performance of \method across various datasets with differing levels of temporal dynamics and emphasize the critical role of community judgment in achieving effective privacy-preserving graph synthesis.

\begin{figure*}[!t]
    \centering
    \includegraphics[width=0.75\textwidth]{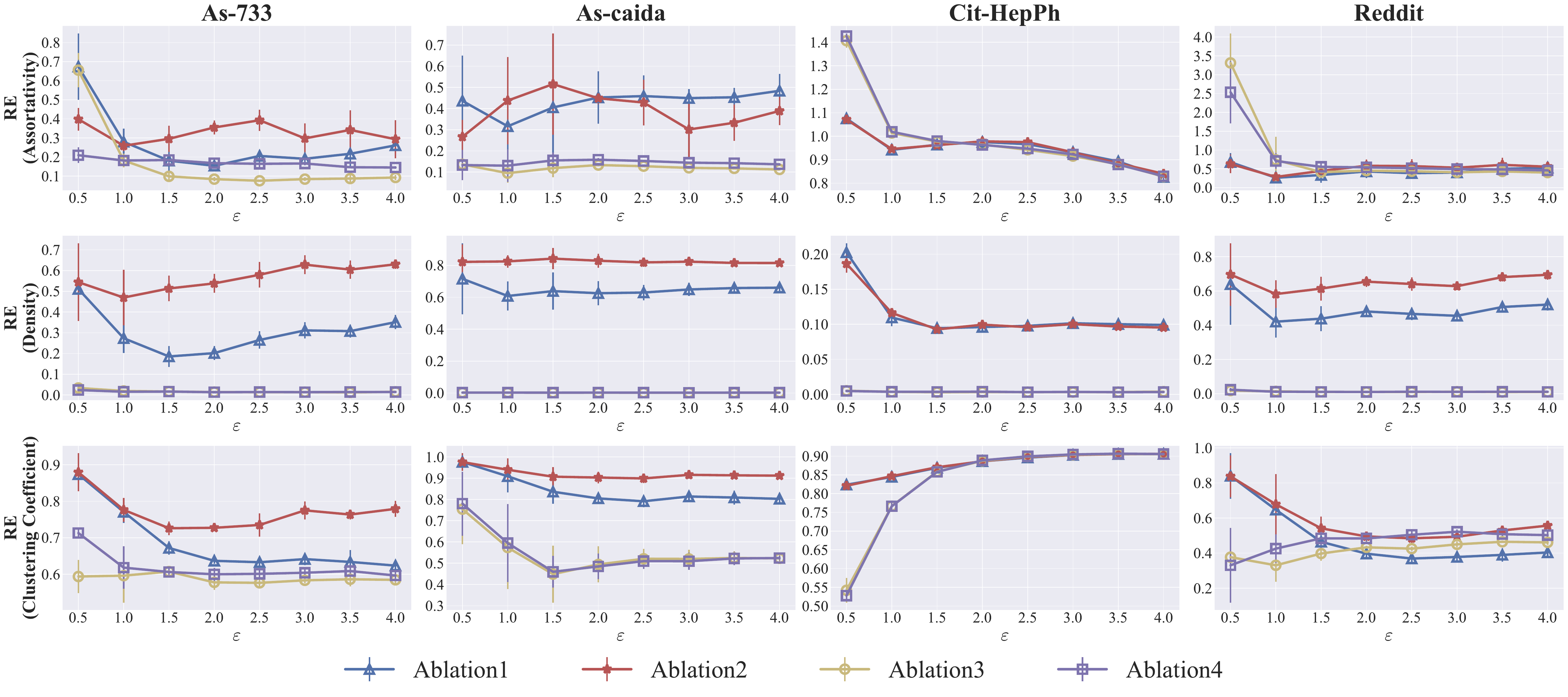}
    \vspace{-0.4cm}
    \caption{
    [Lower is better] Effectiveness of various components.
    Ablation1 represents \method without information fusion and post-processing, Ablation2 represents \method without post-processing,
    Ablation3 represents \method without information fusion,
    and Ablation4 represents \method.
    The columns represent the used datasets, and the rows stand for different metrics.
    In each plot, the $x$-axis denotes the privacy budget $\varepsilon$ and the $y$-axis denotes performance.
    }
    \label{fig:ablation_component_3_metric}
    \vspace{-0.2cm}
\end{figure*}

\mypara{Impact of Various Components}
\autoref{fig:ablation_component_3_metric} illustrates the impact of various components on three metrics.
Ablation2, which only incorporates information fusion, exhibits poorer performance on these metrics. 
This is primarily because the inclusion of information from previous moments may disrupt the total counts of edges and triangles in the current state.
Thus, the REs of Ablation2 in density and clustering coefficients are higher compared to Ablation1 without information fusion, especially when the privacy budget is high.
Ablation3 and Ablation4, incorporating post-processing, demonstrate better performance in density and clustering coefficient metrics than Ablation1 and Ablation2. 
This improvement arises from the constraints imposed on total edge count and node degree information by post-processing techniques.
For the assortativity coefficient,
when the privacy budget is small,
Ablation3 without information fusion is less effective than Ablation4 with information fusion,
As the privacy budget increases, the disturbance at the current timestamp decreases, leading to a notable improvement in the effectiveness of Ablation3.

\section{Parameter Variation}
\label{subsec:appendix_parameter_variation}

\begin{figure*}[!t]
    \centering
    \includegraphics[width=0.75\textwidth]{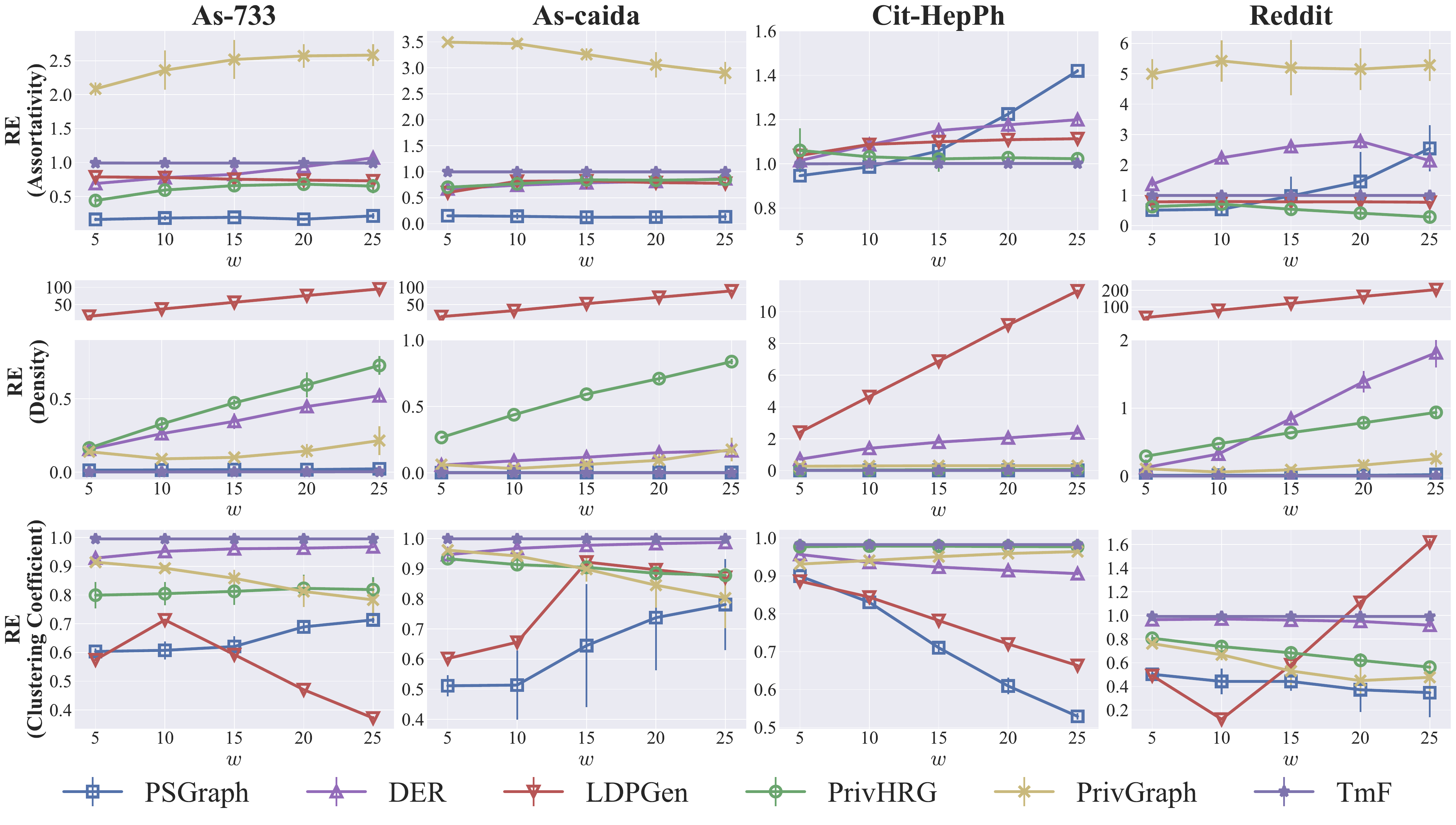}
    \vspace{-0.4cm}
    \caption{
    [Lower is better]
    Impact of window size. 
    The columns represent the used datasets, and the rows stand for different metrics.
    In each plot, the $x$-axis denotes the number of window size $w$ and the $y$-axis denotes performance.
    }
    \label{fig:vary_w_3_metric}
    \vspace{-0.3cm}
\end{figure*}

\mypara{Impact of Window Size}
\autoref{fig:vary_w_3_metric} illustrates the results for three metrics across different $w$ when $\varepsilon=2.5$.
Overall, \method demonstrates competitive performance across most scenarios. 
Regarding the assortativity coefficient, \method outperforms other methods on the first two datasets but shows increasing REs on the latter two datasets as $w$ increases. 
This is mainly due to pronounced variations in the Cit-HepPh dataset where the dynamic mechanism of \method has limited impact under small privacy budgets.
Additionally, the assortativity coefficient of Reddit is low, and the noise in degree information extracted by \method with large $w$ is significant, resulting in high REs.

For density, both \method and \tmf exhibit performance close to 0, 
while other methods show significant increases with $w$, underscoring the importance of the post-processing step for \method. 
In terms of clustering coefficient, \method and \gen perform better across different $w$ values compared to other methods. 
However, the trend in clustering coefficient varies across datasets as $w$ increases, influenced by the varying clustering situations. 
For instance, for the Cit-HepPh dataset with high clustering, \method and \gen tend to produce more triangles as $w$ increases, which in turn leads to smaller REs.

\begin{figure*}[!t]
    \centering
    \includegraphics[width=0.75\textwidth]{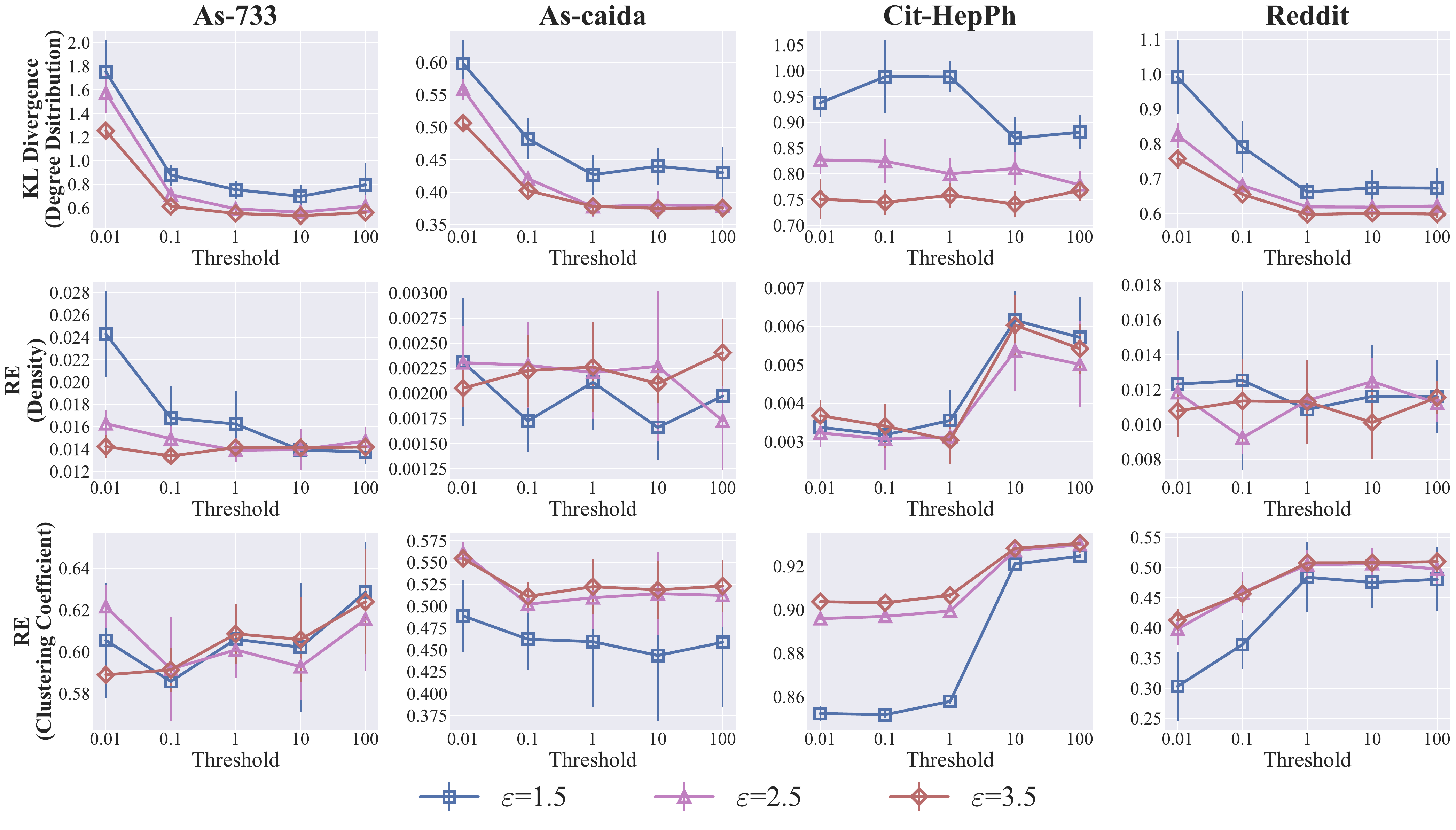}
    \vspace{-0.4cm}
    \caption{
    [Lower is better]
    Impact of threshold. 
    The columns represent the used datasets, and the rows stand for different metrics.
    In each plot, the $x$-axis denotes the threshold and the $y$-axis denotes performance.
    }
    \label{fig:vary_threshold_3_metric}
\end{figure*}

\mypara{Impact of Threshold}
\autoref{fig:vary_threshold_3_metric} illustrates the performance of three metrics across various thresholds. 
Similar to~~\autoref{fig:vary_threshold_2_metric}, the choice of threshold is influenced by the dataset, privacy budget, and metric type. 
\method achieves favorable results with KL divergence at larger thresholds because such thresholds effectively preserve similar information across different timestamps, thereby enhancing accuracy in degree distribution.
For density, \method exhibits larger errors on the Cit-HepPh dataset at larger thresholds due to drastic inherent time variations. 
For clustering coefficients across most datasets, \method shows smaller errors at smaller thresholds, which contrasts with the trend observed in KL divergence.
Consequently, there is no universally optimal threshold that satisfies all configurations.
When the threshold is set to the number of nodes,  \method can acheive competitive results across varied settings.

\section{Detailed Results}
\label{subsec:appendix_detailed_analysis}

\begin{figure*}[!t]
    \centering
    \includegraphics[width=0.75\textwidth] {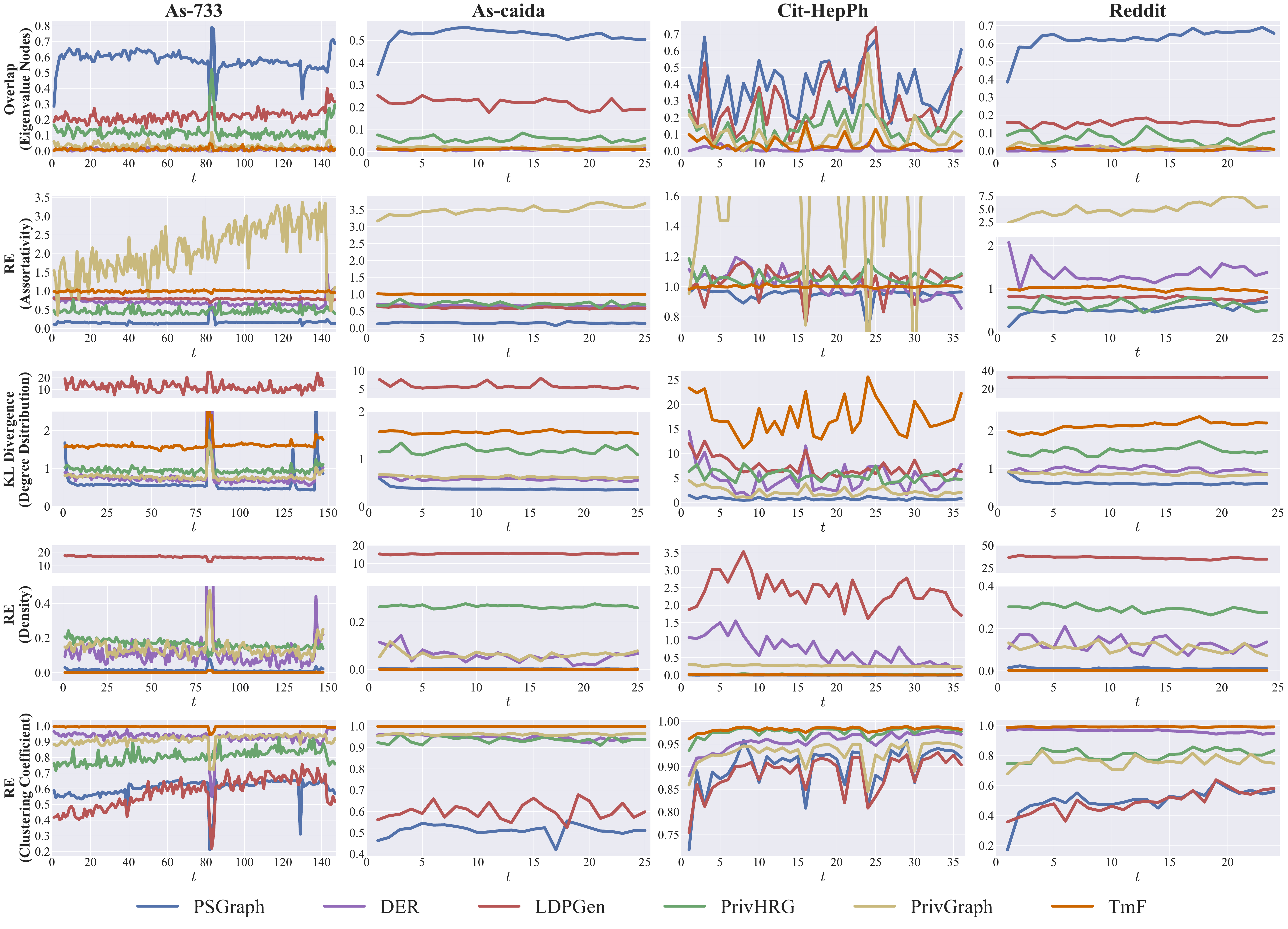}
    \vspace{-0.4cm}
    \caption{
    Performance of various methods at different timestamps.
    The columns represent the used datasets and the rows stand for different metrics. 
    In each plot, the $x$-axis denotes the timestamp, and the $y$-axis denotes the performance. 
    For the first row, higher is better.
    For the last four rows, lower is better.
    }
    \vspace{-0.2cm}
    \label{fig:ti_all_dataset}
\end{figure*}

In~\autoref{fig:ti_all_dataset}, we further provide the performance of all methods at various timestamps when the privacy budget is $2.5$, allowing for a more comprehensive analysis.
\method demonstrates superior performance across nearly all metrics. %
This highlights the critical role played by the dynamic community determination mechanism, temporal information utilization,
and post-processing steps of PSGraph in preserving the intrinsic
characteristics of streaming graphs.
In addition, notable variations are observed in the Cit-HepPh dataset due to its lower temporal dynamics.
\gen performs well in eigenvalue node and clustering coefficient, leveraging $k$-means clustering. 
Yet, it exhibits larger errors in density and KL divergence, attributed to potential edge over-generation under low privacy budgets. 
\privg focuses on protecting the edges within the community while neglecting the edges of nodes outside the community, resulting in low KL divergence in degree distribution and larger REs in the assortativity coefficient. 
\tmf achieves a density close to the original graph, as it takes this into account in the method design. 
\hrg utilizes hierarchical random structures to estimate node-edge probabilities, effectively preserving eigenvalue nodes. 
For \der, random edge assignment leads to significant errors in dense edge graphs.

\section{Ethical Use of Data}
\label{subsec:ethical_use_data}
We strictly followed ethical guidelines by using publicly available, open-source datasets under licenses permitting research and educational use.

\end{document}

%% file: mydefs.tex
\usepackage{epsfig,endnotes,xspace,url,diagbox}
\usepackage{amsmath,amsthm,amsfonts}

\usepackage{enumerate}
\usepackage{varioref}
\usepackage{graphicx}

\usepackage{caption}
\usepackage{subcaption}
\usepackage{float}
\usepackage{epstopdf}
\usepackage{xcolor}
\usepackage{multirow}
\usepackage{comment}
\usepackage{color}
\usepackage[utf8]{inputenc}
\usepackage{mathtools}
\usepackage{wrapfig}
\usepackage{tabularx}
\usepackage{xspace}
\usepackage[noend,ruled,linesnumbered]{algorithm2e}

\usepackage{cases}
\usepackage{stfloats}
\usepackage{fancyhdr}
\usepackage{booktabs}

\usepackage{enumitem}
\setlist[itemize]{leftmargin=*}

\hypersetup{
  colorlinks,
  linkcolor={blue!70!green},
  citecolor={green!70!blue},
  urlcolor={orange!70!red}
}

\newif\ifcomment
\commenttrue

\newcommand{\etal}{\textit{et al.}\xspace}
\newcommand{\ie}{\textit{i.e.}\xspace}
\newcommand{\eg}{\textit{e.g.}\xspace}

\newcommand{\etc}{\textit{etc.}\xspace}

\newcommand{\mypara}[1]{\noindent\textbf{#1.} \xspace}

\renewcommand{\Pr}[1]{\ensuremath{\mathsf{Pr}\left[#1\right]}\xspace}

\newtheorem{theorem}{Theorem}

\newtheorem{definition}{Definition}

\definecolor{revision}{RGB}{0,0,0}
\newcommand{\rev}[1]{{\color{revision} #1\xspace}\xspace}
\newcommand{\revstart}{\begin{color}{revision}}
\newcommand{\revend}{~\!\!\end{color}}

\newcommand{\method}{\ensuremath{\mathsf{PSGraph}}\xspace}
\newcommand{\hrg}{\ensuremath{\mathsf{PrivHRG}}\xspace}
\newcommand{\der}{\ensuremath{\mathsf{DER}}\xspace}
\newcommand{\tmf}{\ensuremath{\mathsf{TmF}}\xspace}
\newcommand{\gen}{\ensuremath{\mathsf{LDPGen}}\xspace}
\newcommand{\privg}{\ensuremath{\mathsf{PrivGraph}}\xspace}

\usepackage{etoolbox}
\makeatletter
\patchcmd{\hyper@makecurrent}{%
    \ifx\Hy@param\Hy@chapterstring
        \let\Hy@param\Hy@chapapp
    \fi
}{%
    \iftoggle{inappendix}{%
        \@checkappendixparam{chapter}%
        \@checkappendixparam{section}%
        \@checkappendixparam{subsection}%
        \@checkappendixparam{subsubsection}%
        \@checkappendixparam{paragraph}%
        \@checkappendixparam{subparagraph}%
    }{}%
}{}{\errmessage{failed to patch}}

\newcommand*{\@checkappendixparam}[1]{%
    \def\@checkappendixparamtmp{#1}%
    \ifx\Hy@param\@checkappendixparamtmp
        \let\Hy@param\Hy@appendixstring
    \fi
}
\makeatletter

\newtoggle{inappendix}
\togglefalse{inappendix}

\apptocmd{\appendix}{\toggletrue{inappendix}}{}{\errmessage{failed to patch}}